\documentclass{emulateapj}

\usepackage{natbib, graphicx}
\newcommand{\kms}{km~s$^{-1}$}
\newcommand{\msun}{$M_{\odot}$}

\newcommand{\feh}{[Fe/H]}

\input epsf

\begin{document}

\title{Chemical Evolution in Hierarchical Models of Cosmic Structure II: The Formation of the Milky Way Stellar Halo and the Distribution of the Oldest Stars}
\author{Jason Tumlinson}
\affil{Space Telescope Science Institute, 3700 San Martin Drive, Baltimore, MD 21218}
\begin{abstract}
This paper presents theoretical star formation and chemical enrichment histories for the stellar halo of the Milky Way based on new chemodynamical modeling. The goal of this study is to assess the extent to which metal-poor stars in the halo reflect the star formation conditions that occurred in halo progenitor galaxies at high redshift, before and during the epoch of reionization. Simple prescriptions that translate dark-matter halo mass into baryonic gas budgets and star formation histories yield models that resemble the observed Milky Way halo in its total stellar mass, metallicity distribution, and the luminosity function and chemical enrichment of dwarf satellite galaxies. These model halos in turn allow an exploration of how the  populations of interest for probing the epoch of reionization are distributed in physical and phase space, and of how they are related to lower-redshift populations of the same metallicity. The fraction of stars dating from before a particular time or redshift depends strongly on radius within the galaxy, reflecting the ``inside-out" growth of cold-dark-matter halos, and on metallicity, reflecting the general trend toward higher metallicity at later times. These results suggest that efforts to discover stars from $z > 6 - 10$ should select for stars with [Fe/H] $\lesssim -3$ and favor stars on more tightly bound orbits in the stellar halo, where the majority are from $z > 10$ and $15-40$\% are from $z > 15$. The oldest, most metal-poor stars -- those most likely to reveal the chemical abundances of the first stars -- are most common in the very center of the Galaxy's halo: they are in the bulge, but not of the bulge. These models have several implications for the larger project of constraining the properties of the first stars and galaxies using data from the local Universe. 
\end{abstract}

\section{Introduction and Motivation}

How and when the first stars and galaxies formed is a central and abiding question in modern astrophysics. The high-redshift frontier has steadily advanced into the end of reionization, $z \sim 6 - 8$, but the truly first stars lie still beyond our view. It is now recognized, on the one hand, that if the first stars are massive and isolated in their own small dark matter halos \citep{Abel:02:93, Bromm:02:23} they will be very faint and difficult to detect directly at high redshift. On the other hand, chemical abundances in the most metal-poor stars may preserve a record of early chemical evolution dating all the way back to the first stars \citep{Freeman:02:487a, Beers:05:531}. This effort, sometimes termed ``Galactic Archaeology'', has the strategic goal of using the most metal-poor stars in the Milky Way and Local Group to address open questions about the first stars and galaxies as a complement to direct study at high redshift.

The goal of this series of papers is to advance the theoretical basis on which these tests can be performed. A large portion of this theoretical foundation consists in demonstrating that the oldest stars in the Milky Way can be used in such a fashion. Stated another way, it would be fatal to the whole enterprise if it turned out that the Milky Way contained no stars from $z > 6$, or from the Epoch of Reionization. As yet we have no purely empirical basis for claiming that it does; all such statements are still necessarily model-dependent. Dark-matter only numerical simulations \citep{White:00:327, Scannapieco:06:285, Diemand:07:859} have already demonstrated that Milky-Way sized DM halos should contain significant structures that originated before $z > 6$ (roughly the current observational frontier). This is well-established and seems necessarily true in the CDM paradigm. Yet while the dark matter holds some interest in its own right, it is still only the dark scaffolding on which the Galaxy's stellar populations are built. When we consider the stars, a host of other questions arise: How many low-metallicity stars should we expect in the Milky Way? How many of these should date from $z > 6$, if any? What are the most advantageous places to look for these true survivors, and what should we expect to find there? To what extent are these early extremely metal-poor (EMP) stars obscured by later populations of similar metallicity but of less relevance to the first stars? All these questions must be answered to accomplish the overarching goal of ``Galactic Archaeology'', but they have no easy answers. 

That metal-poor stars\footnote{This paper adopts the \cite{Beers:05:531} terminology for labeling metal-poor stars, e.g. EMPs, or extremely metal-poor stars, have [Fe/H] $<-3$, and so on.} address the first stars has been taken for granted as a working hypothesis by many observational programs \citep{Cayrel:04:1117, Christlieb:02:904, Frebel:05:871}. Yet this claim rests on no firm empirical proof: though some metal-poor stars are known to have ages consistent with 13 Gyr ago, these ages cannot yet be determined precisely or for a large sample. The observational approach rests rather on the somewhat different claim that these stars are related to the first stars by way of the ``chemical clock''; they are considered related to the first star formation in their region of space by virtue of their low metallicity, not by their age. Yet not all [Fe/H] $\sim -3$ stars are necessarily created equal, and some may arise much later and be chronologically unrelated to the epoch of first star and galaxy formation in which we are interested. The relationship between the two clocks is likely to be a complex one. Further, even if we take the relationship between EMP stars and the first stars at face value, this assumed relationship does not tell us directly how much and what kind of information EMPs contain about the first stars, or how many stars are needed to answer the open questions. 

These issues of interpretation reduce to a core problem that must be solved for the larger enterprise to advance. That is: the readily observable quantities in the present-day Galaxy -- the luminosity, temperature, chemical abundances, and orbits of individual stars and their trends in stellar populations -- are only indirectly related to the properties of stars and gas in early galaxies that preceded the Milky Way, and to the physical processes that governed their formation. For example, nucleosynthetic yields from supernovae enter newly formed stars only after some degree of dispersal and mixing in interstellar gas. The degree of mixing may vary with time and place. For another example, long-lived stars formed in the shallow potential wells of early galaxies -- the exact populations we now study in the MW halo -- have been stripped from their parent halo and had their present orbits in the Galaxy set by a long and stochastic series of mergers and interactions that cannot be reconstructed exactly, if at all. These two examples are likely essential ingredients in the chain of events leading from the stars we want to understand to the stars we actually observe. There appears to be little alternative to an attempt at a single model that includes these processes in building quantitative links between theoretical ideas about early galaxies and observations in the present day. 

The observational frontier is being advanced by large surveys that detect and measure galactic structure and substructure, and by mining these surveys for metal-poor stars and then measuring their detailed abundances in intensive spectroscopic campaigns. As examples of the former, there is now a proliferating number of known stellar streams \citep{Yanny:00:825} and faint dwarf galaxies (Belokurov et al. 2007) discovered in the Sloan Digital Sky Survey. Using SDSS, \cite{Carollo:07:1020} confirmed two distinct chemical and kinematic components of the stellar halo near the Sun, using a much larger sample than available before SDSS. As examples of the latter, the HK  \citep{Beers:92:1987} and HES \citep{Cohen:04:1107} surveys have been thoroughly mined for metal-poor stars, leading to discovery of the most metal-poor stars known \citep{Christlieb:02:904, Frebel:05:871}, to the discovery of a class of carbon-enhanced metal-poor stars (Lucatello et al. 2005), to increasingly precise measurements of the stellar metallicity distribution \citep{Schoerck:08:1172}, and to intensive studies of the abundances and distribution of the heaviest chemical elements \citep{Barklem:05:129}. These discoveries are important motivations for this theoretical work. They provide crucial information about the structure of the Milky Way's halo for which any full synthetic model must ultimately account. Each of these individual studies provides data on a key piece of the overall puzzle; collectively, they suggest that a very large quantity of information may lie in stellar halo populations. Yet observational studies have a strong incentive to carefully select their target populations, so it is difficult to draw general inferences about the behavior of the full Milky Way system without a model that calculates its predictions at a level of detail commensurate with the observations. Given the complexity of the datasets and the large dynamic range between the extremes of the Milky Way on the one hand and abundances in individual stars on the other, building such model presents a significant challenge. 

Paper I in this series \citep{Tumlinson:06:1} made an attempt at such a model in the form of a new stochastic, hierarchical model of Galactic chemical evolution, based purely on the ``Extended Press-Schecter'' merger-tree formalism \citep{Somerville:99:1}. That model favored close tracking of chemical abundances over a realistic dynamical model of halo formation, with the goal of constraining the IMF of the first stars by way of their contributions to chemical abundances at low metallicity. That technique worked adequately for the tests performed in that paper, which tracked gas and chemical enrichment budgets with a high level of detail but which could not calculate the spatial or kinematic distributions of model stellar populations. By itself, it yields only a part of a model. Using a similar semi-analytic merger-tree technique, \cite{Salvadori:07:647} studied the chemical evolution of the halo in response to the first stars, but with the similar limitation that no detailed treatment of the dynamics was included. At the same time, \citeauthor{Bullock:05:931} and their collaborators built dynamical models of halo formation based in non-cosmological N-body simulations and applied simple chemical evolution prescriptions to assess the relationships between kinematics and chemical abundances \citep{Robertson:05:872, Font:06:585, Font:06:886}. Their work has been compared extensively to surveys of the Milky Way halo and has been shown to provide a realistic description of the halo's basic properties - mass, degree of substructure, and trends in chemical abundances. While each was successful in their own way, none of these studies excelled at understanding the formation of the first stars and galaxies by locating their descendants in the Milky Way. Paper I lacked any treatment of dynamics, and the \citeauthor{Bullock:05:931} halo models did not simulate the Milky Way in a cosmological ``live halo'' and did not follow the merger histories of the dwarf galaxies or -- most importantly -- the formation history of the Milky Way host. Because of these limitations neither the Paper I nor the \citeauthor{Bullock:05:931}  models were well-suited for studying the present-day properties and distribution of stellar populations formed at very high redshift. 

The need for a strong synthetic and flexible Milky Way model and the limitations of the previous efforts inspire the present study, which attempts to retain the strengths of the previous work while rectifying some of its major weaknesses. With respect to Paper I, the present paper takes a step backwards to reassess a number of key assumptions that underlie the basic approach to probing the early Universe with local signatures; the spirit of the new approach is to develop a workable model with a minimum degree of complexity. These reassessments are possible because the present paper deals with a much more sophisticated realization of the Galaxy's dark matter halo, and they are necessary because the earlier study made a number of simplifying assumptions that are now testable in detail. The goals of the present study are: 
 
\begin{itemize}
\item[1.] To assess the fundamental basis for the ``Galactic Archaeology'' effort \citep{Beers:05:531} by demonstrating that the Milky Way should contain stellar populations that date from the Dark Ages and the Epoch of Reionization, by extending previous DM-only modeling efforts to account for how the Galaxy's progenitors acquire and process their gas and build up their chemical abundances. 
\item[2.] To identify the most important global and local physical influences on the chemodynamical assembly of the Milky Way's stellar halo, to assess the sensitivity of the resulting stellar populations to these influences, and to help guide future in-depth studies.
\item[3.] To lay out a framework for the calculation of key observational quantities related to Galactic chemical evolution, in absolute physical units where possible, and their sensitivity to the key influences; for example, how does cosmic reionization affect the radial density profile of stellar populations in the halo? 
\item[4.] To suggest methods by which stellar populations that originate at high redshift can be isolated from later populations with which they may be easily confused in random samples; for example, when looking at any random star with [Fe/H] $< -2$, what is the chance that it reflects conditions at $z > 6$?
\end{itemize} 

It is with all these ideas in mind that this paper is organized as follows:  Section~2 describes the N-body simulations of Milky-Way-like DM halos and \S~\ref{section-results1} describes the mass assembly history of the inner portions of the MW halo, considering dark matter only. Section~\ref{section-methods2} describes the chemical evolution models that run within the dark-matter halo merger trees. Section~\ref{section-results2} shows how the models provide a realistic description for the gross properties of the real Milky Way halo, as an observational check on the model construction and parameter choices. Section~\ref{section-results3} shows the metal abundances of stars from the first MW progenitors and maps out broad trends for their variation within the MW halo at the present time. Section~\ref{section-results4} discusses the general results of the models and draws some implications for observational studies of metal-poor stars and star formation in the early Universe. 

\section{Methods I: N-body Halo Simulations}

\begin{figure*}[ht]
\begin{center}
\plotone{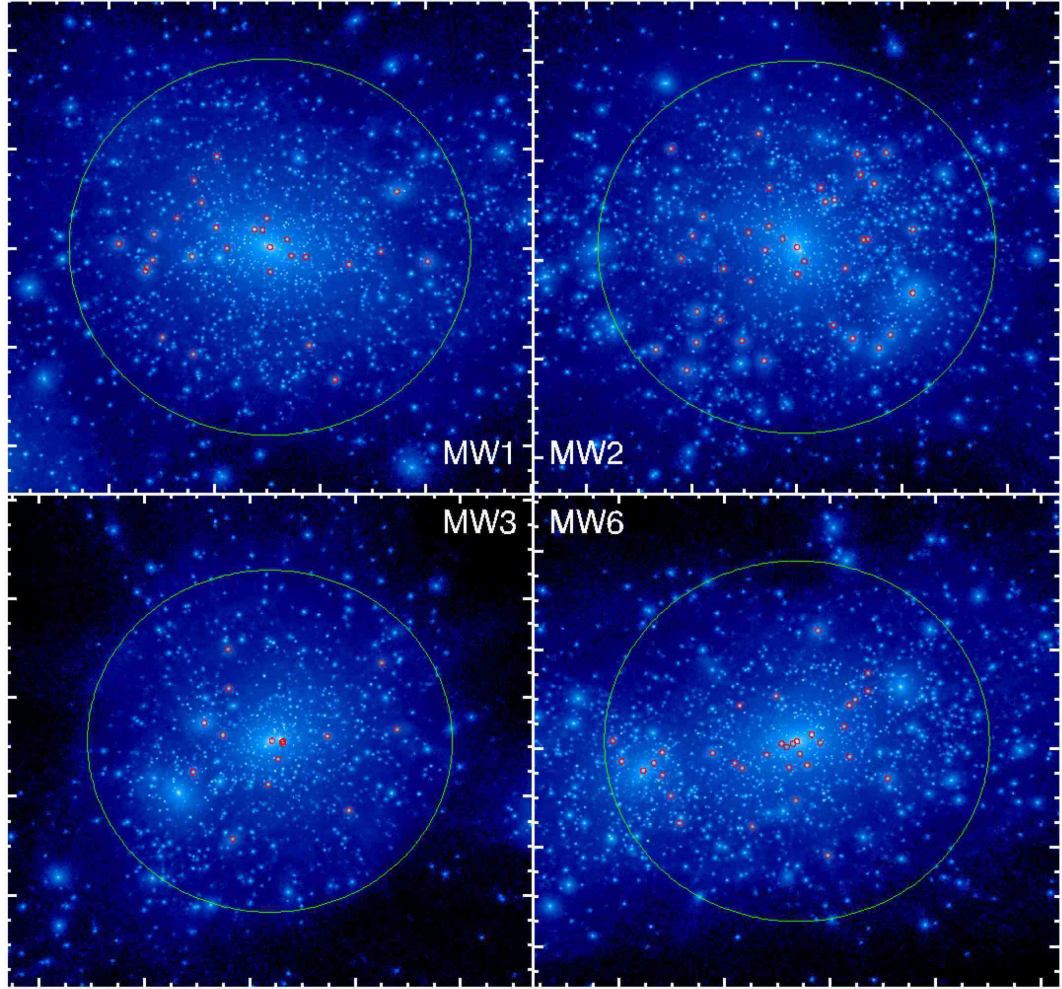}
\caption{Projected view of the dark-matter density field for the four MW-like halos at $z = 0$. The virial radius is marked with the large circle, and subhalos with $v_c \geq 20$ \kms\ and $R \leq R_{vir}$ are marked with small circles (the host included). \label{viewplot}} 
\end{center}
\end{figure*}

\subsection{Simulation Setup and Initial Conditions} 

The first step in constructing a full chemodynamical model is to simulate the formation of a dark-matter halo that resembles the Milky Way in its important properties. This section describes the construction of the Milky-Way-like halo runs, and their analysis up through the construction of merger trees. 

The N-body simulations were run using the Gadget2 code (version 2.0; \citealp{Springel:05:1105}) in parallel mode on a local computing cluster. The initial conditions were created using the GRAFIC1 initial conditions package \citep{Bertschinger:01:1}. These initial conditions are specified for a WMAP3 cosmology (3-year mean; see Tables 2 and 4 of \citealp{Spergel:07:377}), with matter density $\Omega _m = 0.238$, baryon density $\Omega _b = 0.0416$, vacuum energy density $\Omega _{\Lambda} = 0.762$, Hubble constant $H_0 = 73.2$ km s$^{-1}$ Mpc$^{-1}$, power spectrum slope $n_s = 0.958$, power spectrum normalization $\sigma _8 = 0.761$, and primordial helium abundance $Y_{\rm He} = 0.249$. GRAFIC was fed a WMAP3 power spectrum calculated for these parameters by CMBFAST \citep{Seljak:96:437}. The output of GRAFIC1 is a set of positions and velocities for cold dark-matter particles that represent a random realization of the density field in a periodic cubic box of comoving size 7.320 $h^{-1}$ Mpc in one dimension (10000 kpc in proper coordinates at $z = 0$). The runs analyzed here began with initial conditions calculated on a grid of 1024 particles in each dimension, which were then binned to obtain initial conditions of size $512^3$ or $256^3$, as desired. At the default $512^3$ resolution, the CDM particles have mass $M_p = 2.64 \times 10^5$ \msun. 

\subsection{Milky Way Analogue Selection and Refinement}

Initial conditions that lead to the formation of a MW-like halo at $z = 0$ were generated as follows. First, 10 random realizations of CDM structure were run with the raw $1024^3$ initial conditions binned to $128^3$ (1 particle for every 512 particles in the raw initial conditions). These runs evolve to $z = 0$ very quickly and are easily searched for MW-like halos for resimulation at the higher resolution. Interesting halos are chosen to have virial masses $M_{200} \approx 1.5  \times 10^{12}$ \msun\ and are subject to the additional constraint that they must not experience a major (3:1) merger after approximately $z = 1.5 - 2$. Because there is only one Milky Way to study observationally, the focus of these simulations has been on studying multiple halos at around the same mass, to look for stochastic fluctuation in merger history, rather than to examine the dependence of properties on host halo mass. 

From the $128^3$ runs four MW-like halos were obtained to run at $512^3$; their properties are listed in Table~\ref{halo-catalog}. Snapshots of the particle positions and velocities were generated at 20 Myr intervals before $z = 4$ and 75 Myr intervals from $z = 0 - 4$, for a total of 236 snapshots for each run. Only a portion of each box centered on the halo of interest and typically $4-6$ Mpc on a side, was resimulated at the higher resolution, while the remainder of the box was run at effective $256^3$ to save time. The gravitational smoothing length for all simulations was 100 pc in comoving coordinates. 

\subsection{Halo finding and characterization}

The particles in the simulation snapshots have only six physically independent quantities to be studied: three components of position and three components of velocity. Each particle is assigned additional derived physical quantities: a local gravitational potential and matter density, based on smoothing the discrete particles over a 32-particle kernel using SKID \citep{Stadel:01}, as well as kinetic energy and angular momentum. 

With completed N-body runs in hand, the next step is to locate and characterize bound halos and their evolution. This ``halo-finding'' step is crucial, since the resulting merger trees are only as good as the underlying halo catalogs on which they are based. A combination of halo finding and characterization techniques provides a good compromise between accuracy and flexibility. It is essential to capture the small halos embedded within the larger potential well of the host halo as it forms. The 6-dimensional friends-of-friends (6DFOF) halo finding technique presented by \cite[][Appendix A2]{Diemand:06:1} meets both requirements effectively. This algorithm was implemented by the author within the original Washington N-body shop FOF software\footnote{http://www-hpcc.astro.washington.edu/tools/fof.html} by adding a velocity term, $b_v$, to the phase-space ``linking length'', so that particles are linked together if their relative positions and velocities satisfy the condition: 
\begin{equation}
\frac{| \vec{x}_1 - \vec{x}_2 |}{b dx} + \frac{| \vec{v}_2 - \vec{v}_2 |}{b_v} < 1 
\end{equation}
where $b$ is the traditional friends-of-friends spatial linking length, $dx$ is the mean particle spacing, and $b_v$ is the velocity-space linking length added for 6DFOF. 

\begin{deluxetable}{ccccccc}[!ht]
\tablecolumns{7} 
\tablenum{1} 
\tablewidth{0pt} 
\tablecaption{Catalog of Dark-Matter Halo Runs} 
\tablehead{ 
\colhead{Name} 
&\colhead{$M_{FOF}$}
&\colhead{$R_{200}$} 
&\colhead{$M_{200}$} 
&\colhead{$c$}
&\colhead{$z_{LMM}$} \\
\colhead{}
&\colhead{$10^{12} M_{\odot}$}  
&\colhead{kpc} 
&\colhead{$10^{12} M_{\odot}$} 
&\colhead{} &\colhead{}   
} 
\startdata 
MW1   &  1.46   & 381 & 1.63   & 12.2 & 2.1 \\
MW2   &  1.50   & 378 & 1.59   &   9.2 & 3.5 \\
MW3   &  1.20   & 347 & 1.23   & 15.5 & 2.0  \\
MW6   &  1.39   & 366 & 1.44   & 13.6 & 3.0
\enddata 
\label{halo-catalog}
\end{deluxetable} 

No single pair of values for $b$ and $b_v$ finds all identifiable halos and accurately characterizes their properties at all redshifts. At high redshift where the typical bound halo is small and highly clustered near like halos but substructure is minimal at the resolution of these calculations, the 6DFOF parameters should converge to the plain FOF parameters, $b = 0.2$, so $b_v \rightarrow \infty$. At low redshift, when most of the halos of interest are subhalos of the host, smaller spatial and velocity-space linking lengths are needed to avoid overlinking of substructure with the host and with other substructure. As $z \rightarrow 0$, it turns out that $b = 0.07$ and $b_v = 40$ km s$^{-1}$ lead to good results when compared with manual fits to the velocity profiles of obvious and isolated substructure. This value of $b$ is used until $z = 5$, and then linearly increases to $b = 0.2$ at $z = 12$, while $b_v$ linearly increases as $160(1+z)/3$ from $b_v = 40$ at $z= 0$. These values are pragmatic compromises that are considered valid only because they work well; the halo catalogs are tested against pure friends-of-friends at high $z$ and found to be complete and accurate above the resolution limit, while at low $z$ the maximum circular velocities of subhalos are well characterized (compared with values calculated for isolated subhalos directly from the particle data) even if the outer regions of small halos are not linked with the cores owing to the small linking lengths. The outcome of this process is a catalog for each snapshot output that contains halo position, maximum circular velocity, and a group identification for each particle. Some calculations use halo maximum circular velocity $v_{c}$, while others use halo masses estimated by direct fits to the velocity profiles or, occasionally, an analytic relation between $v_{c}$ and $M_{halo}$. Alternate halo catalogs and merger trees were derived for all snapshots using a pure FOF algorithm (that is, $b=0.2$, $b_v \rightarrow \infty$ for all $z$) for use in tests where accurate masses for large halos are desired and accurate measures of substructure are not. 

\subsection{Merger Tree Construction and Particle Tagging}

Merger trees are constructed by linking together halos in successive snapshot outputs according to rules that determine whether an earlier halo in fact enters a later halo. In merger tree parlance, each halo is a ``node'' in the tree; the earlier halo is known as the ``child'' halo, while the later halo in time is known as the ``parent''. Working from high redshift, the algorithm links together pairs where the candidate child has more than 50\% of its particles in the parent. Some fraction of halos becomes unbound by, e.g. gravitational interaction with the much larger host halo. These branches of the tree terminate in a ``child'' halo with no parent that is marked with a special flag as a disrupted subhalo. These particles that become unbound are typically found in a larger tree node at a later snapshot. This merger tree mechanism makes it easy to study the mass assembly history of the host halo and its substructure. This process also generates a large two-dimensional particle-halo cross reference table that specifies the node in the tree where each particle was located (if it was bound) at each redshift snapshot. This cross-reference table makes it easy to correlate particle properties at $z = 0$ with their history of accretion and merging throughout the merger tree.  
\begin{figure}[!ht]
\begin{center}
\plotone{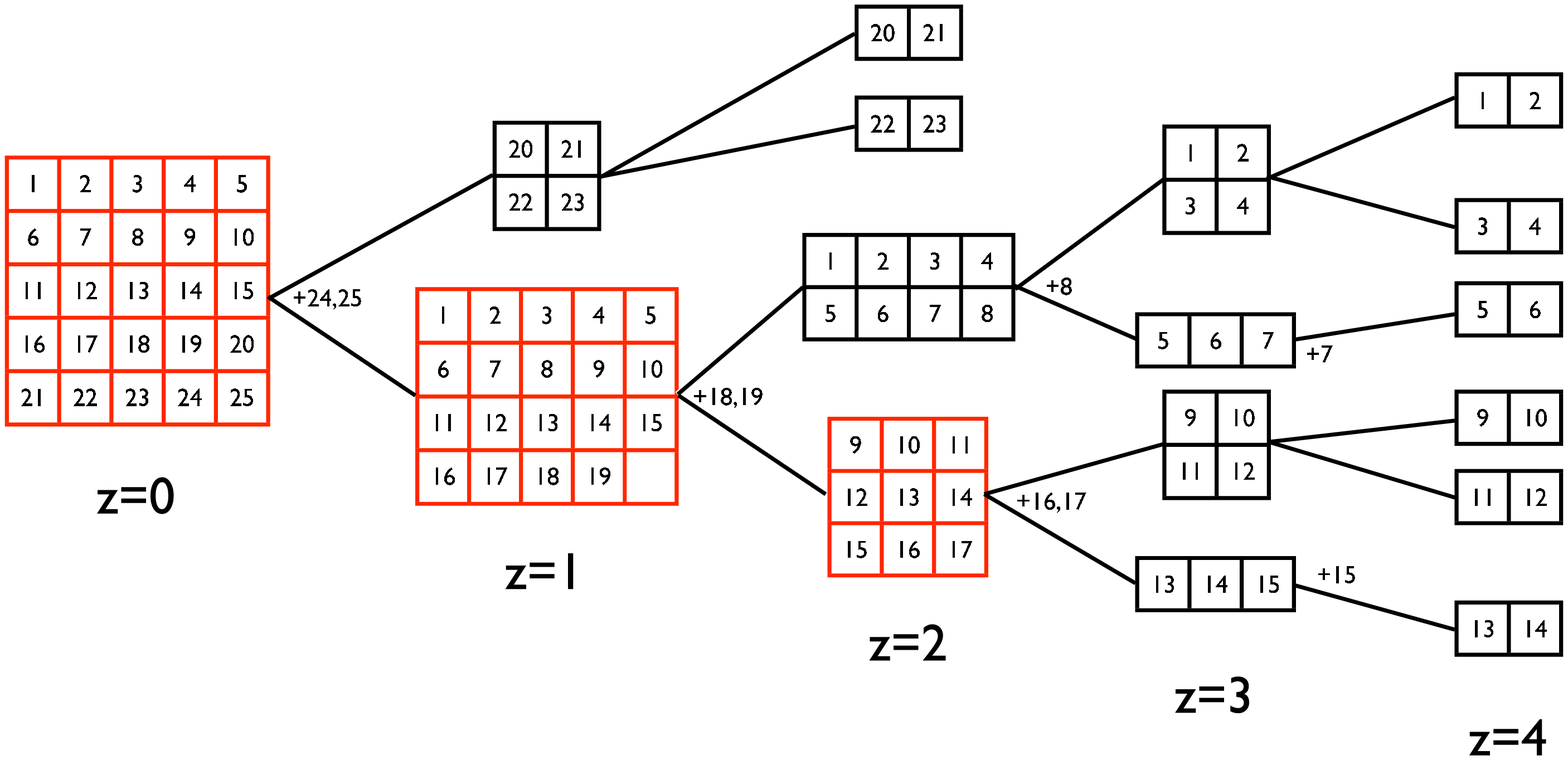}
\caption{Schematic merger tree showing mass acquisition by mergers and accretion. The three large halos at $z = 0$, 1, and 2 are schematically tagged as the host. See \S~\ref{tree-explain-section} for discussion.} \label{tree_example_plot} \end{center}
\end{figure}

\section{Results I: Mass Assembly Histories of the Milky Way Halo and Surviving Subhalos} 
\label{section-results1}

\subsection{Basics}
\label{tree-explain-section}

Dark-matter only simulation of cosmic structure is by now a mature technique that has generated many useful insights into the dynamical assembly of the Galaxy's dark-matter halo \citep{Helmi:03:834} and its satellites \citep{Zentner:03:49, Kravtsov:04:609}. The reader is referred particularly to the series of papers on the {\it Via Lactea} simulations \citep{Diemand:07:859} and their successors for deep insight into the development of these simulations and their implications for dark matter substructure. The goal of these paper not to advance the frontier of  the simulations themselves but to obtain measures of halo assembly that will illuminate the expected distribution of halo stellar populations, particularly those arising before reionization. The first step in such a study is to analyze how particles enter and move through the merger tree. The goal of this effort is to assess how and when the dark halo acquires its mass, and how this behavior might influence the resulting stellar populations. This analysis is illustrated with a merger-tree schematic in Figure~\ref{tree_example_plot}. 

First, each DM particle is assigned the redshift when it first entered any bound halo according to the group-finding algorithm. This ``redshift of first entry'', $z_{entry}$, is used as an approximate measure of when the particle could first be associated with stars. In the example tree, particles $1-6$, $9-14$, and $20-23$ enter new bound halos from the unbound state, and $z_{entry}$ for these groups is 4, 4, and 2, respectively. Particles 7, 8, $15-19$, and $24-25$ enter bound halos that already existed at an earlier redshift. Their $z_{entry}$ is, e.g. 3 for particle 15 and 1 for particles 18 and 19. A useful modification of this is to assign $z_{entry}$ only when the particle has entered the densest or most bound portion of a halo - particles that are loosely bound are easily stripped, and perhaps should not be associated with stars since they do not trace the inner, tightly bound portions of a DM halo where stars form. This modified version of $z_{entry}$ is used here, with the condition that the particle must be among the 10\% most bound particles in its halo. This quantity can also be considered proxies for the age of the oldest (metal-poor) stars that could be carried by the particle.

\begin{figure*}[!ht]
\begin{center}
\plotone{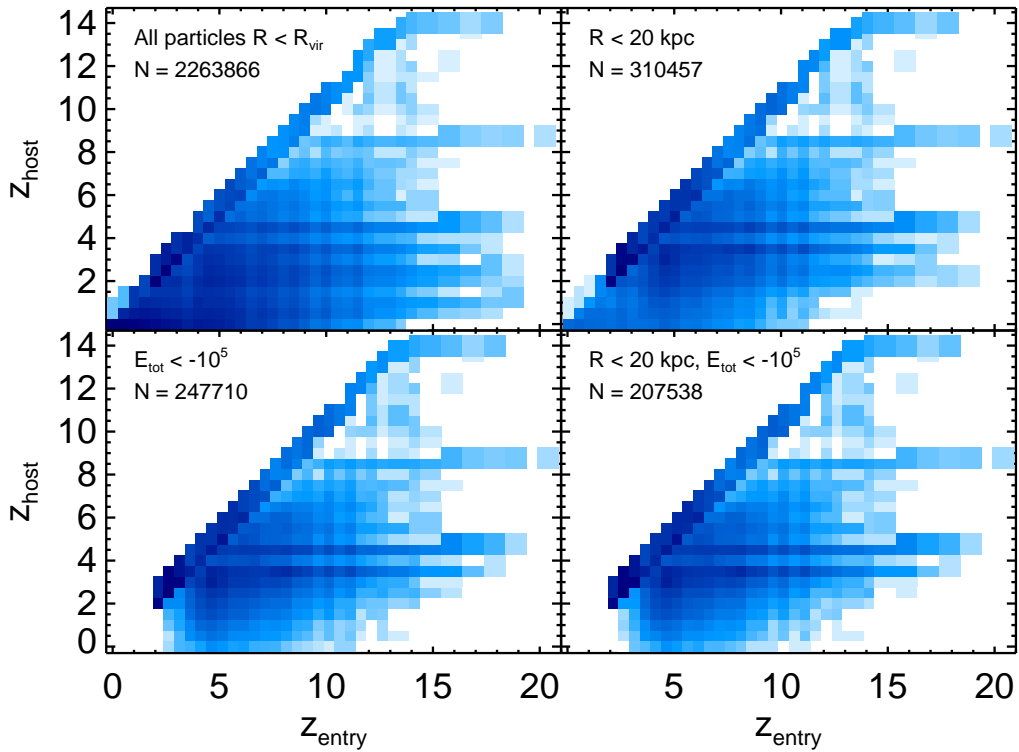}
\caption{Two-dimensional histograms for the redshift of first entry $z_{entry}$ and redshift of host entry $z_{host}$ for MW1, four different cuts on particle radius or binding energy, and excluding those particles that accrete directly into the host after the last major merger. This analysis uses the ``pure FOF'' trees and counts particles as bound only when they have entered the most bound 10\% of halo's mass.} \label{zplot-figure} \end{center}
\end{figure*}

Each particle is also assigned $z_{host}$,  the redshift when the particle enters the host halo. The identity of the host halo is determined by working backwards recursively from $z = 0$ and at each redshift choosing the most massive progenitor. The ``redshift of host entry'', or $z_{host}$, is always later, and typically much later, than $z_{entry}$ for the same particle. This distinction between $z_{entry}$ and $z_{host}$ reflects the basic hierarchical nature of structure formation, in which small subhalos form and then accrete into larger pieces over time. 

A third type of particle behavior in the tree is important for characterizing the phase-space structure of the central regions of the host halo. At late times, $z \lesssim 2$, when the host halo has attained a total mass of $\sim 10^{11}$ \msun, it rapidly accretes matter in particles that have never previously been in a bound halo. Generally the host halo (as marked in Figure~\ref{tree_example_plot}) is tracked out to the redshift of the last major merger ($z_{LMM}$), and particles that accrete directly into the host at this redshift or later are given a special tag as ``accreted particles'' that enter through smooth accretion into the host. In the example, particles $16-19$ and $24-25$ are late-accreted particles and are assigned this flag. In many comparisons below, these particles are excluded from phase-space density and structure comparisons because they should not be associated with stars. Of course, any particle that accreted from an unbound state into the host, but was previously bound to and then stripped from a bound halo at an earlier redshift, will not receive this special tag and will be counted as eligible to carry stars.

\subsection{Mass Assembly History as Expressed by Particles} 
\label{mass-section}

These simple tools can now be used to analyze the mass assembly history of the MW analogue halos. Figure~\ref{zplot-figure} shows a probability density plot for selected subsets of particles from MW1, using the ``pure FOF'' (\S~2.3) trees and assigning $z_{entry}$ only when particles that have entered the inner 10\% of a bound halo. The color code expresses the relative number of stars in each square bin with black representing the most populated bin in each panel. The upper left panel shows all particles within the virial radius, including those in identifiable subhalos. When all particles are considered, the construction of the halo as expressed by $z_{host}$ extends all the way to the present and many particles that entered bound objects very early enter the host only relatively late, virtually always as the inner, tightly bound portions of an identifiable subhalo. Considered as a whole, the Galactic DM halo grows continuously at all times, up through the present. 

When only the inner regions of the halo, within R = 20 kpc of the center of mass, are considered, the picture changes substantially. Now the most populated bins are centered around $z_{entry} \sim 5$, $z_{host} \sim 2-4$. In this panel the effect of large subhalos is plain: they appear as dark horizontal stripes since they build up individually over a range of $z_{entry}$ and enter the host together at a single $z_{host}$. Comparing this panel with the first, it appears that the inner portions of the halo form early with respect to the whole halo. 

With a cut on particle binding energy (lower left), the early assembly of the inner halo is even more evident. Now the low-redshift tail of $z_{entry}$ disappears. These are particles that lie within $R \leq 20$ kpc but whose orbital energies place them on trajectories that are either unbound (in  a few cases) or which extend far out into the outer regions of the halo and which happen to lie in the 20 kpc sphere at $z = 0$. The overall ``center of gravity'' in the figure is however, little changed, since the inner 20 kpc sphere is composed mostly of tightly bound particles. This is clearly demonstrated by the lower right panel, which imposes both cuts but is nearly indistinguishable from the panel with the $E_{tot}$ cut alone. This exercise reveals the twofold essence of the hierarchical picture of galaxy formation; first, that large subcomponents of galaxies form independently and merge together later, and second, that the inner, more tightly bound material formed and accreted earlier than outer, more loosely bound material. Simply put, halos form hierarchically and from the inside out. 

\begin{figure*}
\begin{center}
\plotone{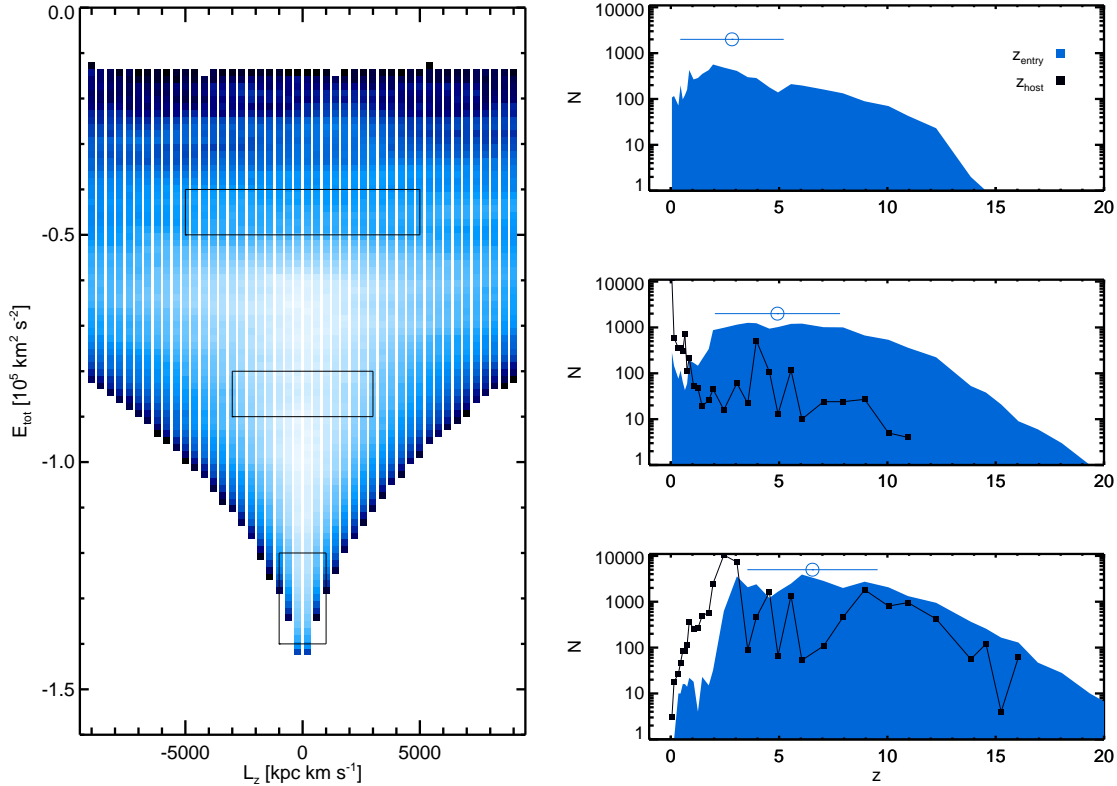}
\caption{Left panel: For MW1, the distribution of particle kinematics plotted as $E_{tot} = E_{kin} + E_{pot}$ vs. the $z$-component of the angular momentum. The origin of the coordinate system lies at the center of mass of the particles within 10 kpc of the densest particle in the host. The black rectangles mark the three regions of interest for assessing the mass assembly history. Right panels: The distribution of two measures of halo assembly for the three regions of interest in the Lindblad plot, from top to bottom. The filled curves mark the distribution of $z_{entry}$. The solid squares mark the distribution of $z_{host}$. See text of \S~\ref{mass-section} for discussion.  \label{mhist-plot}} \end{center}
\end{figure*}

Figure~\ref{mhist-plot} examines these results in more depth by showing how $z_{entry}$ and $z_{host}$ relate to binding energy in detail. 
The right panel is a Lindblad plot of MW1. Three regions of $L_z$-$E_{tot}$ space are extracted and the distribution of $z_{entry}$ and $z_{host}$ are plotted at right. Particles that accrete directly to the host after $z_{LMM}$ are excluded, as are particles that are in identifiable subhalos at $z = 0$. With $z_{entry}$ as the proxy (blue regions), the typical particle that ends up in the lower two regions enters a bound halo at $z \gtrsim 5$, while the less bound material does not enter until significantly later. The open circles with range bars mark the mass-weighted mean and 1$\sigma$ variance for $z_{entry}$ for these three regions. The trend is the same here; more bound particles entered earlier, with half them in bound objects before $z = 6$. The most bound subregion also possesses a longer tail to $z > 10$, which indicates that the very first DM halo antecedents of the Milky Way formed very early indeed. 

The distribution of $z_{host}$ shows an even stronger trend with binding energy; it is clear from comparing the black lines in the two lower panels that the most bound portions of the halo formed early, with peaks corresponding to major mergers culminating in the last one at $z = 2.1$, after which there is a steep decline in mass being added on these tightly bound orbits (particles acquired by ``smooth accretion" are excluded here). While the middle region receives material from these same high-redshift mergers, it also continues to receive significant material from a series of minor mergers up until $z=0$, including some subhalos that may have their own histories extending back to $z > 5$. Much of this material is in identifiable streams. 

From Figures~\ref{zplot-figure} and \ref{mhist-plot} it is clear that although the particles that build the inner regions of the halo join the host at around $z_{LMM}$, they have been in bound subhalos for much longer, $z = 4 - 20$. For region 1, the least bound, the mean redshift of first entry is $\langle z_{entry} \rangle = 3$, while for the most tightly bound subregion 3, $\langle z_{entry} \rangle = 6$ (see range bars in the figure). At high redshift, this generally means formation into a newly formed halo or accretion into an existing one, but at $z = 6$ the host halo (the most massive halo that ends up in the final $z = 0$ halo) is only one of dozens of subhalos that will eventually enter the host. Thus the inner dark-matter halo of the Milky Way is assembled from many pieces that begin their individual histories very early, and likely contain stars that formed over a wide range of redshift back to the earliest epochs. The important implications of these results for stellar populations will be taken up in later sections of the paper (\S~\ref{section-results2}). 

There is one important proviso to these results: because these simulations have finite mass resolution, the redshift of first entry for each particle is strictly a lower limit. That is, a higher-resolution simulation might find that a given parcel of mass entered a bound object at an earlier redshift but that the bound object in question is too small to be resolved by the simulations (which counts only halos with $\geq 32$ particles). However, since $M_p = 2.6 \times 10^5$ \msun\ and $32M_p = 8 \times 10^6$ \msun, all but the smallest progenitor halos at the highest redshifts are already resolved by these simulations. Thus resolution effects should not severely bias these results, and if they do the mean redshifts and oldest stellar ages are pushed to earlier times, and all subregions in $E_{tot}$ will systematically shift together. 

One of the chief goals of this work is to assess the extent to which stellar populations in the Milky Way can be used to explore the conditions for galaxy and star formation during the epoch of the first galaxies. The chief conclusion of this investigation into the merger and assembly histories of MW-like dark matter halos is that there is good reason to expect that the MW itself contains stars formed at $z > 6 - 10$, and that {\it this conclusion holds independently of how or whether baryonic gas physics are modeled}, provided small halos at high redshift are able to form any stars at all. In summary, this study of MW-like dark-matter halos and their merger tree histories leads to the following conclusions:
\begin{itemize}
\item[1.] Though the inner region of the halo near the Sun as a structure forms ``late'' (around the time of the LMM), approximately half of the particles were in place at $z > 6$.
\item[2.] Halo formation is intrinsically hierarchical, in the sense that pieces form and are then assembled into larger structures. If any aspect of chemical evolution depends on the mass of the halo hosting star formation (i.e.~depth of local potential well), a model must account for the full hierarchy of mergers and its development over time. 
\item[3.] Evidently the properties of stellar populations in the inner regions of the halo will depend on details of how baryons behave in small dark-matter halos at $z > 4$. This is of course a topic of interest in itself, with as yet little observational leverage. Since it appears that Milky Way halo stellar populations will be sensitive to dwarf galaxies at high redshift, these populations can be used to answer open questions about star formation in the early Universe more generally. This will be taken up in \S~6. \end{itemize}


\section{Methods II: Modeling of Gas Processes and Stellar Populations} 
\label{section-methods2} 

\subsection{Introduction to the General Approach}

The previous two sections studied the mass assembly history of MW-like halos using dark-matter only simulations. For these models to compare to real observations of Galactic structure it is necessary to translate them somehow into model stellar populations. This is what the Paper I method was designed to do: to implement Galactic chemical evolution calculations in a generic, hierarchical merger tree representation of the Galaxy. With merger trees drawn from N-body simulations, this technique becomes even more powerful, since it is now possible to track the phase-space motions of individual particles and the stellar populations associated with them. The general approach will be to assume that each DM halo found in the simulations, and followed through the merger tree, possesses some gas that has been accreted from the IGM, that this gas forms stars, that these stars return metals and energy to the host halo and to the larger environment, and that future generations of stars form with the now metal-enriched gas. These are all processes that can and have been followed with simple analytic relations as in many past models of chemical evolution. The difference here is that the gas budgets, star formation rates, and feedback processes can be written in terms of, or related back to, the underlying dark-matter scaffolding; while this is technique has a history of application to problems of galaxy formation at large, in cosmological volumes, its application to single galaxies, with full attention paid to detailed chemical evolution, is relatively new \citep[BJ05; Paper I;][]{Robertson:05:872, Font:06:585}. 

When attempting this kind of modeling it is desirable to control and understand the influence of the parameters describing the behavior of the baryons, and to avoid excessively complicated parameterizations. While it is true that the differential equations that control ``star formation'' and ``chemical evolution'' in these models describe the underlying physical processes only indirectly, it is not true that the parameterization of our ignorance by establishing these relations then permits any desired solution to arise. Proceeding from the simplest possible relations and testing against observations at each step can identify those quantities, such as gas budget and luminosity, that have close relationships, and those that do not, such as interstellar mixing and reionization. As shown below, even simple descriptions of gas budgets, star formation efficiency, and mass return from small halos provide a workable and informative model of Galactic chemical evolution in the hierarchical context. 

Paper I presented a new chemical evolution code that works both within the hierarchical context of galaxy formation and in the stochastic limit of low-metallicity Galactic evolution.  In that paper the hierarchical galaxy assembly history was specified by the common technique of halo merger trees \cite{Somerville:99:1} to decompose the Galaxy into its precursor halos working backward in time. It then calculates the history of star formation and chemical enrichment in these objects working forward in time, keeping track of all individual metal-producing supernovae and assigning metallicity to new star formation stochastically from all prior generations. The second generation of the chemical evolution code offers several improvements over the version used in Paper I. These improvements are described here in detail. The parameter choices adopted for fiducial models appear in Table 2. 

\subsection{Ingredients of Gas and Star Modeling}

\subsubsection{Baryon Assignment}
\label{baryonassignmentsubsection}

\renewcommand{\arraystretch}{1.3}
\begin{deluxetable}{ccl}[t]
\tablenum{2}
\tablecolumns{6} \tablewidth{0pc} \tablecaption{Model Parameters and Their Fiducial Values}
\tablehead{Parameter & Value & Description }
\startdata
\cutinhead{Cosmology}
$\Omega _m$ & 0.238 & Matter density \\
$\Omega _b$  & 0.0416 & Baryon density \\
$\Omega _\Lambda$ & 0.762 & Vacuum energy density \\
$H_0 $    & 73.2 & Hubble constant [km s$^{-1}$ Mpc$^{-1}$] \\ 
$n_s$      & 0.958 & power spectrum slope \\
$\sigma _8$ & 0.761 & power spectrum normalization \\ 
$Y_{He}$ & 0.249 & primordial Helium abundance\\
\cutinhead{Chemical Evolution}
$f_{bary}$           & 0.05          & baryonic mass fraction \\ 
$\epsilon_{*} \equiv 1/\tau _*$ & $1 \times 10^{-10}$ & Star formation efficiency [yr$^{-1}$] \\
$m_{\rm Fe}^{II}$ & $0.07-0.1$ & SN II iron yield [\msun] \\
$f_{Ia}$           & 0.015          & SN Ia probability \\ 
$m_{\rm Fe}^{Ia}$  & 0.5       & SN Ia iron yield [\msun]  \\
$f_{esc}^{Z}$   &  50      & Escape factor of metals \\
$\epsilon_{SN}$ & 0.0015 & SNe energy coupling 
\enddata
\label{parameter-table}
\end{deluxetable}

Knowing how much gas a galaxy contains is the first step to knowing how many stars it can form, so the assignment of baryonic mass budgets to dark-matter halos is the crucial first step in modeling their chemical evolution. The term ``assignment'' is used purposely to connote a rule for baryonic accretion, not a rigorous physical model. The most straightforward such rule simply assigns each halo a gas budget $M_{gas}$ such that $M_{gas} / M_{halo} = \Omega _{b} / \Omega _{m}$, which takes the value $0.175$ in the adopted cosmology. Because it does not take into account the effects of internal or external feedback influences on the gas budget, particularly reionization, this prescription is not a very useful one.

Instead, these models adopt a prescription based on that of \cite{Bullock:05:931} that takes into account heuristically the influence of a photoionizing background from the aggregate star formation in all galaxies. This very simple model assigns a fixed mass fraction of baryons, $f_{bary}$ to all DM halos before reionization, $z_{r}$. After $z_r$, gas accretion and therefore star formation in small halos are suppressed completely below $v_c = 30$ \kms. Between 30 \kms\ and 50 \kms, the assigned baryon fraction varies linearly from 0 to $f_{bary}$.  

This baryon assignment prescription is the minimal one that can be effective at suppressing gas accretion and star formation by low mass halos. It is intended to capture the IGM ``filtering mass'' (Gnedin 2000) below which halos are too small to retain baryons that have been heated to $T \gtrsim 10^4$ K by global reionization. More recent determinations of this filtering mass (Okamoto et al. 2008) have reduced it somewhat from Gnedin's original formulation, but the prescription adopted here already permits gas accretion into halos at $M_h \sim 10^9$ \msun\ after reionization. The exact minimum mass for baryon accretion should be regarded as uncertain; testing of adopted values for the minimum $v_c$ that can accrete baryons after reionization, down to 15 \kms, indicates that the results for the assembly of the halo are not too sensitive to this adjustment (see \S~5.1). 

\subsubsection{Star Formation Efficiency} 

Stars are formed in discrete ``parcels'' with a constant efficiency, $\epsilon_*$, such that the mass formed into stars $M_* = \epsilon_* M_{gas} \Delta t$ in time interval $\Delta t$. The star formation efficiency is equivalent to a timescale, $\epsilon _* = 1 / t_*$, on which baryons are converted into stars. The fiducial choice for this parameter is $t_* = 10$ Gyr, or $\epsilon _* = 10^{-10}$ yr$^{-1}$. 

\subsubsection{Stellar Initial Mass Function}

Paper I devoted significant effort to exploring how the initial mass function of the first stars could be constrained using chemical abundances in EMP stars. IMF variations in the EMP populations themselves are indicated by the relative fraction of carbon-enhanced metal-poor stars (CEMPs) in the halo \citep{Lucatello:05:833, Tumlinson:07:1361, Tumlinson:07:L63}. Refining and strengthening this technique, and applying it to future surveys of EMP stars, is an important goal of this theoretical effort. However, variations in the IMF of the first stars have relatively little effect on the bulk properties of the halo, and IMF variations more generally deserve their own intensive study. Here the IMF is assumed to be invariant at all times and at all metallicities, to simplify the parameter space. IMF variations will be studied carefully once other influences such as gas accretion, star formation, and feedback are better understood in the context of a global model. The invariant IMF adopted here is that of \citet[][eq. 1 and 2]{Kroupa:01:231}, $dn/dM \propto (m/M_{\odot})^\alpha$,  with slope $\alpha = -2.3$ from $0.5  - 140$ \msun\ and slope $\alpha = -1.3$ from $0.1 - 0.5$ \msun.

\subsubsection{Type Ia SNe}

To model the chemical evolution of heavy elements for longer than $\sim 100$ Myr it is necessary to include the yields of Type Ia SNe. Type Ia SNe are assumed to arise from thermonuclear explosions triggered by the collapse of a C/O white dwarf precursor that has slowly accreted mass from a binary companion until it exceeds the 1.4 \msun\ Chandrasekhar limit. For stars that evolve into white dwarfs as binaries, the SN occurs after a time delay from formation that is roughly equal to the lifetime of the least massive companion, which much evolve off the main sequence before the larger star, now a white dwarf, can accrete its outer envelope, exceed the critical mass, and explode.

This process is captured stochastically as follows. Stars with initial mass $M = 1.5 - 8$ \msun\ are considered eligible to eventually yield a Type Ia SN. When stars in this mass range are formed, some fraction of them, $f_{Ia}$, are assigned status as a Type Ia and given a binary companion with mass ratio compared to the pair mass $\mu$. To account for the tendency of the mass ratio to approach 1:1, the binary companion mass is chosen from the probability distribution, 
\begin{equation} 
P(\mu) = 2^{1+\gamma} (1+\gamma) \mu^{\gamma} 
\end{equation} 
where $0 \le \mu \leq \frac{1}{2}$ and $\gamma = 2$  \citep{Greggio:83:217}. The prospective SN Ia is then assigned an explosion time corresponding to the lifetime of the less massive companion. 

The chemical evolution results are not sensitive to the choice of $\gamma$, but they do depend on the SN Ia probability normalization, $f_{Ia}$. This parameter is fixed by normalizing to the observed relative rates of Type II and Type Ia SNe for spiral galaxies in the local universe \citep{Tammann:94:487}, such that a fraction $f_{Ia} = 0.015$ of all stars formed with $1.5 - 8$ \msun\ eventually experience a Type I SN. This normalization gives a ratio of SN II to Ia of 6 to 1. This parameter also varies with the IMF slope and with the mass limits assumed to lead to a Type II SNe. The adopted choice of $f_{Ia}$ was calculated for a long, steady star formation history going back 10 Gyr, to resemble the SFH of galactic disks, and for the \cite{Kroupa:01:231} IMF and $10 - 40$ \msun\ for Type II SNe.

\subsubsection{Chemical Yields} 

Typical yields for Type II and Type Ia supernovae are drawn from literature sources. Type II supernovae are assumed to arise from stars of $10 - 40$ \msun, with mass yields provided by Nomoto (2006, private communication). They represent the bulk yields of core-collapse supernovae with uniform explosion energy $E = 10^{51}$ erg. These models have $M = 0.07 - 0.15$ \msun\ Fe per event. For Type Ia supernovae with $1.5 - 8$ \msun\ the models adopt the W7 yields of \cite{Nomoto:97:467} for Fe, with 0.5 \msun\ of Fe from each Type Ia SN. Since it is the goal of this study to model the bulk formation of the Galactic halo and to explore kinematic trends, the present models track only bulk metallicity with Fe as the proxy reference element. Relative abundances of other elements and their trends will be studied in a later paper.

\subsubsection{Interstellar gas mixing}

Paper I implemented a scheme for stochastic mixing of supernova yields into the interstellar and intergalactic medium. 
These factors are critical for determining the abundance patterns of individual stars, since the observed abundances are 
a mixing-weighted average of all prior generations. Since the aims of this paper are focused on studying the global formation and kinematics of the MW halo, and not detailed abundances, these models assume instantaneous mixing within the available gas reservoir. 
Numerically, the ``dilution mass'' $M_{dil} = 10^{11}$ \msun, but in practice is limited to the gas budget of the host halo.  The mixing timescale is set to $t_{dil} = 1000$ yr, which is very short compared with the average timestep used in the chemical evolution calculation. These parameters recover the instantaneous mixing limit and are not varied here (they are therefore left out of Table 2). 

\subsubsection{Chemical and Kinematic Feedback} 
\label{feedbacksubsection}

One possible cause of the observed luminosity-metallicity (L-Z) relation for Local Group dwarf galaxies is supernova-driven mass loss from small DM halos~\citep{Dekel:03:1131}. This process is physically complex and must be included in the models in terms of a highly simplified prescription controlled by one or two adjustable parameters. The present models use a prescription that parallels the scheme developed by \cite{Robertson:05:872} and used subsequently with good success by  
\cite{Font:06:585, Font:06:886, Font:08:215} to model the Local Group L-Z relation.  Their work specified the mass loss rate in terms of the instantaneous supernova rate, which is directly proportional to the star formation rate for a given IMF. Since the present model tracks star formation events and therefore supernovae individually, it should track mass loss in terms of the number of supernovae per timestep in a way that takes into account the intrinsic time variability in the star formation rate and rate of supernovae from a stochastically sampled IMF. Intuitively, supernova-driven winds should respond to the depth of the host halo's gravitational potential well, and to the energy imparted to the ejecta by the supernovae. The prescription should also allow for selective loss of gas and metals, following indications that superwinds flowing from nearby galaxies are metal-rich \citep{Low:99:142}. The total SN energy available to drive the wind is given by
\begin{equation}
E_{SNe} = \epsilon _{SN} \sum_{i} N_{SN}^i E_{SN}^i = 10^{51} \epsilon _{SN}\sum_{i} N_{SN}^i E_{51}^i
\end{equation}
where the sum over index $i$ sums over all supernovae from past timesteps that are just undergoing an explosion in the current timestep, and $\epsilon _{SN}$ is the fraction of total supernova energy that is converted to kinetic energy of the remnant. The kinetic energy in the wind is then given by
\begin{equation}
E_{wind} = \frac{1}{2} M_{lost} v_{wind}^2
\end{equation}
where $M_{lost}$ is the total gas mass lost to unbound winds and
$v_{wind}$ the wind velocity.  Energy balance requires that $E_{SNe} = E_{wind}$, so that
\begin{equation}
10^{51} \epsilon _{SN}\sum_{i} N_{SN}^i E_{51}^i = \frac{1}{2} M_{lost} v_{wind}^2
\label{wind-eq1}
\end{equation}

For the wind to escape, $v_{wind}$ must exceed the escape velocity of the parent DM halo, where $v_{escape} = 2 v_{c}$ for a halo with maximum circular velocity $v_{c}$. Substituting this condition into Equation~\ref{wind-eq1} and rearranging: 
\begin{equation}
M_{lost} = \frac{ 10^{51} \epsilon _{SN} \sum_i N_{SN}^i E_{51}^i }{2 v_{virc}^2 }
\end{equation}
Normalizing to $v_{circ} = 50$ km s$^{-1}$ and performing additional substitutions:
\begin{equation}
M_{lost} \simeq 10^4 \epsilon _{SN}\sum_i N_{SN}^i E_{51}^i  \left( \frac {v_{circ}}{50 \,{\rm km\, s}^{-1}} \right)^{-2}
\label{wind-eq2}
\end{equation}
At each timestep, this mass of gas is becomes unbound and is removed permanently from the gas reservoir.  Equation~\ref{wind-eq2} contains only one free parameter, $\epsilon  _{SN}$, which expresses the fraction of the supernova energy that is converted to kinetic energy retained by the wind as it escapes. For instance, if 5\% of the total SN energy is converted to kinetic energy, and 3\% of this is imparted to the ejected material, then $\epsilon _{SN} = 0.0015$. This approach is well-suited to the stochastic framework of Paper I, since it counts individual supernovae and allows for variations in the number and energy of SNe from timestep to timestep. As desired, this expression includes the total energy available from SNe to drive winds and the depth of the DM gravitational potential that the winds must escape. 

The selective loss of metals that should arise when supernovae drive their own ejecta out of the host galaxy is captured by a new parameter $f_{esc}^Z$, which expresses the increased metallicity of the ejected winds with respect to the ambient interstellar medium. At each timestep, a total mass in iron $M_{lost}^{Fe}$ is removed from the gas reservoir of the halo: 
\begin{equation}
M_{lost}^{Fe} = f_{esc}^Z M_{lost} \frac{M_{ISM}^{Fe}}{M_{gas}}
\end{equation}
where $M_{ISM}^{Fe}$ is the total mass of iron in the ambient interstellar medium, $M_{gas} \times 10^{\rm [Fe/H]}$. This prescription ensures that, on average, the ejected winds are $f_{esc}^Z$ times more metal-enriched than the ambient interstellar medium. Alternatively, the fraction of metal mass lost from the halo is $f_{esc}^Z$ times higher than the total fraction of gas mass lost. This behavior, and the scaling with halo circular velocity, are consistent with the treatment by \cite{Robertson:05:872}, with two key differences in formulation. First, this prescription is generalized to write the mass loss in terms of the instantaneous number of supernovae per timestep rather than the instantaneous star formation rate. Second, the typical values of $f_{esc}^Z$ are somewhat lower here; for Robertson's choice of ejection parameters $f_{esc}^Z$ as defined here takes on the value 112 (see their equations $19-21$). As will be shown below, the best fitting value of $f_{esc}^Z$ for these models is somewhat lower, $\sim 50$, because they take into account the full merger tree history of each dark matter halo rather than an average mass assembly history specified by the mass at accretion. 

The adopted prescription is certainly not the only prescription of mass and metal loss driven by supernovae that could be adopted. This one is preferred because it takes into account energy balance and its scaling with potential well depth, because it can work with a stochastic supernova rate, and because it is easily written to include selective loss of metals. Its most important feature is that it works; it gives reasonable results with reasonable physical inputs.

\subsubsection{Isochrones and Synthetic Stellar Populations}

To compare these model halos to observational data on the real Milky Way and its dwarf satellites, it is necessary to calculate the luminosities and colors of model stellar populations using precalculated isochrones and population synthesis models. Each star formation parcel possesses a metallicity, age (with respect to a fiducial time, usually $z = 0$), and a total initial mass distributed according to the assumed IMF. These three quantities together uniquely specify an isochrone and how it is populated. These models adopt the isochrones of \cite{Girardi:02:195} and \cite{Girardi:04:205} for the UBVRIJHK and SDSS $\it ugriz$ systems, respectively, as published on the Padova group website\footnote{http://pleiadi.pd.astro.it/}. Their machine-readable tables provide the absolute magnitudes for each bandpass as a function of initial stellar mass, which are then integrated over the assumed IMF. In this way a total luminosity in each passband is obtained. There is a small adjustment ($\lesssim 0.1$ mag) to correct for the fact that the Padova isochrones terminate at 0.15 \msun\ instead of the assumed lower mass limit of 0.1 \msun. The lowest available metallicity in these isochrones is $Z = 0.0001$, or [Fe/H] $= -2.3$. Isochrones with $Z = 0$ were calculated by \cite{Tumlinson:03:608} and \cite{Marigo:01:152}; the former did not include ages beyond $10^8$ yr and the latter did not include masses below $0.7$ \msun, so neither provides the complete coverage of age and metallicity that is required by the present models. Instead, the models adopt the lowest metallicity from the uniform \cite{Girardi:02:195} set to represent all stellar populations with lower metallicities. This approximation is suitable for the tests performed below, where the luminosity of stellar populations below [Fe/H] $=-2.3$ is small compared with those at higher metallicity. With these isochrones it is also possible to derive colors and luminosities for only parts of a stellar population by, e.g. integrating only down the giant branch, which can facilitate comparisons with specific observational results. 

\subsubsection{Parcels, Batches, and Particle Tagging}

The chemical evolution code tracks gas, stellar mass, and metal abundance budgets through the merger tree. The first step in the calculation is to specify an endpoint or ``root halo'' for which the complete star formation and chemical enrichment history will be determined. In Figure 3 this is the 25-particle z = 0 halo, which is the host in the example. The endpoint may also be a smaller halo that is disrupted by interaction with the host, in which case its tree terminates at $z > 0$. Next, the highest redshift progenitor halo of the endpoint is determined from the merger tree, and the chemical evolution calculation starts there. The star formation history is calculated using a small number of timesteps, usually 10 $-$ 50 between each redshift snapshot. At each timestep, a star formation ``parcel'' is created with a single metallically and IMF. The metallicity for the parcel is derived from the present gas metallicity, as described above. Each parcel is a single-age simple stellar population that will be used later to integrate up the star formation history of the complete halo. For example, if the 17-node tree in Figure 3 were computed with $N_{timesteps} = 10$, the resulting number of parcels would be 170. For each node in the tree, there are $N_{timesteps}$ parcels, and when halos merge their lists of parcels (from which supernovae and yields will be drawn to assign metallicity to later parcels) are concatenated. This process continues until the ``root'' endpoint is reached and all progenitor nodes have been calculated. 

Since the final Milky Way halo contains multiple subhalos that have experienced (more or less) independent star formation histories, multiple endpoints or ``root halos'' must be calculated. More than one endpoint is calculated in ``batches'' that have a single parameter set, and which collectively express the full star formation and chemical evolution history of the full MW halo. 

To explore the spatial, kinematic, and dynamical properties of stellar populations in the Galactic halo it is necessary to assign ``stars'' to DM particles in the N-body simulation. This ``tagging'' step proceeds as follows. For each particle in the simulation, a cross-reference table is constructed that contains an index to the node of the tree in which each particle was located at each snapshot output. For the example tree in Figure~\ref{tree_example_plot}, this table is an array consisting of 25 rows (particles) and 5 columns (redshifts). In the example particle 1 would have five non-zero entries in its column, while particle 20 would have 3 and 24 only one. To obtain a list of stellar populations associated with each particle, these indices are used to obtain the list of star formation parcels that occurred within the nodes in the tree where that particle resided over time. Each particle is assigned a fraction of that parcel in proportion to its fraction of the mass in each halo node in which it was present. In most cases the particles are screened such that the mass of each star formation parcel is associated with only the inner (densest or most bound) 10\% of the particles in each halo node, to approximate the effect of stars forming deep in the potential well of the galaxy and to avoid assigning stars to particles that are only loosely bound to their host. Stars are assigned individually for each parcel, particle, and timestep combination, so complex histories for individual particles are handled correctly. This might occur, for instance, when a particle associated with stars is stripped from a small halo during accretion into a larger halo into an unbound state, only to later accrete into the larger halo. This scheme also makes it easy to study subsets of the stellar populations, say all stars formed before $z = 6$, or with [Fe/H] $<-3$, simply by excluding those regions of the cross-reference table. All the later measures of stellar populations are obtained in this way.

\section{Results II: Observational Checks on the Fiducial Model}
\label{section-results2}

\subsection{Stellar Halo Mass and Density Profile}

\begin{figure}
\hspace{-0.2in} 
\plotone{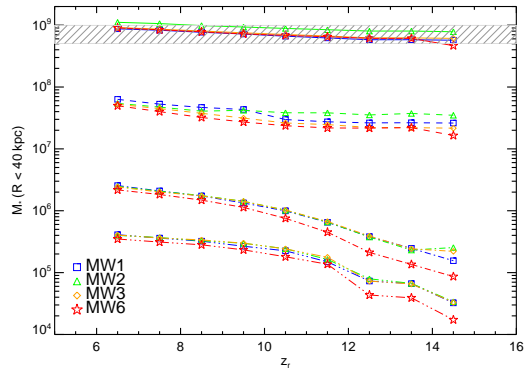}
\caption{ The total stellar mass with $R \leq 40$ kpc for the four model halos and four different cuts on metallicity. The upper set of curves counts all stars of any metallicity, and the three lower curves mark [Fe/H] $ < -2$, $-3$, and $-4$ from top to bottom. For these models $\epsilon _* = 10^{-10}$ and $f_{bary} = 0.05$. The hashed band marks the range inferred from SDSS star counts by \cite{Bell:08:295}, corrected to account for different assumed IMFs. 
\label{hmassfig} }
\end{figure}

\begin{figure*}
\plotone{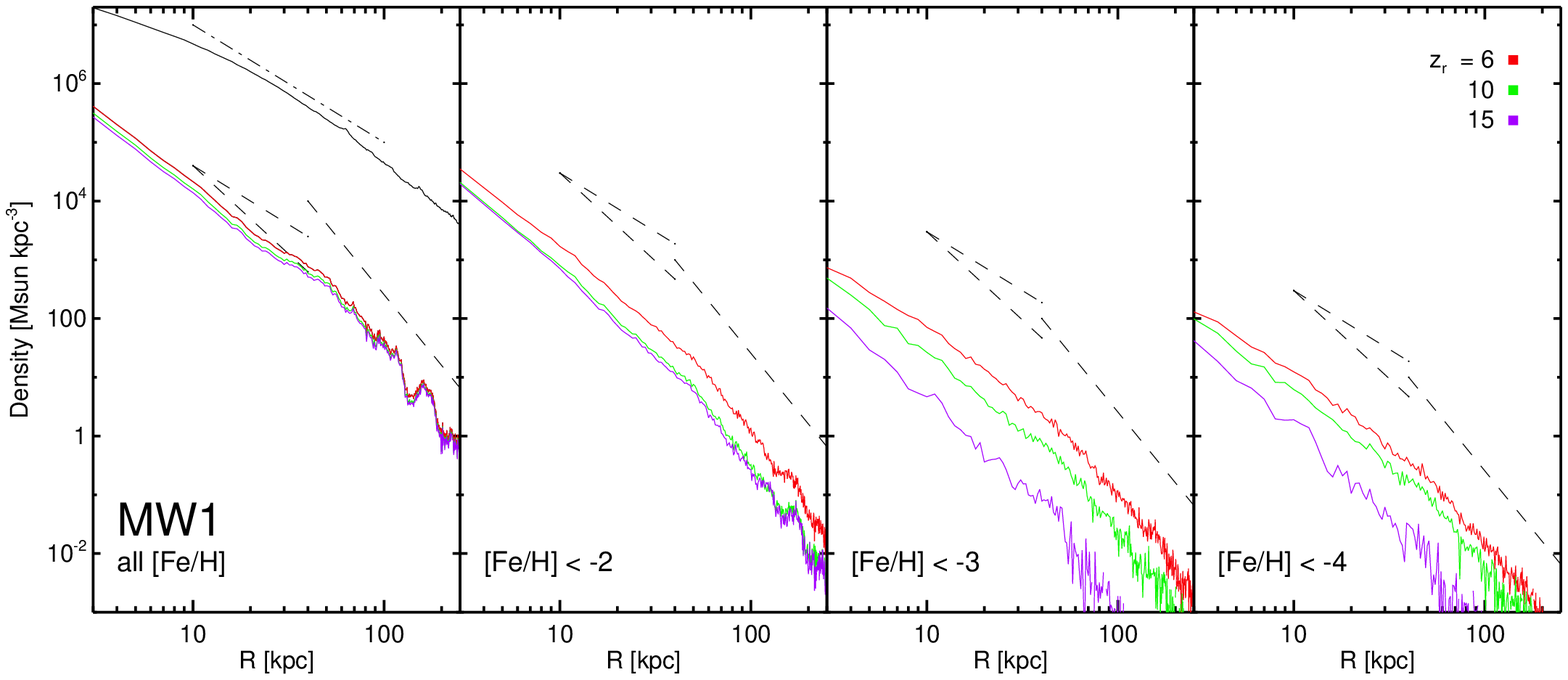}
\plotone{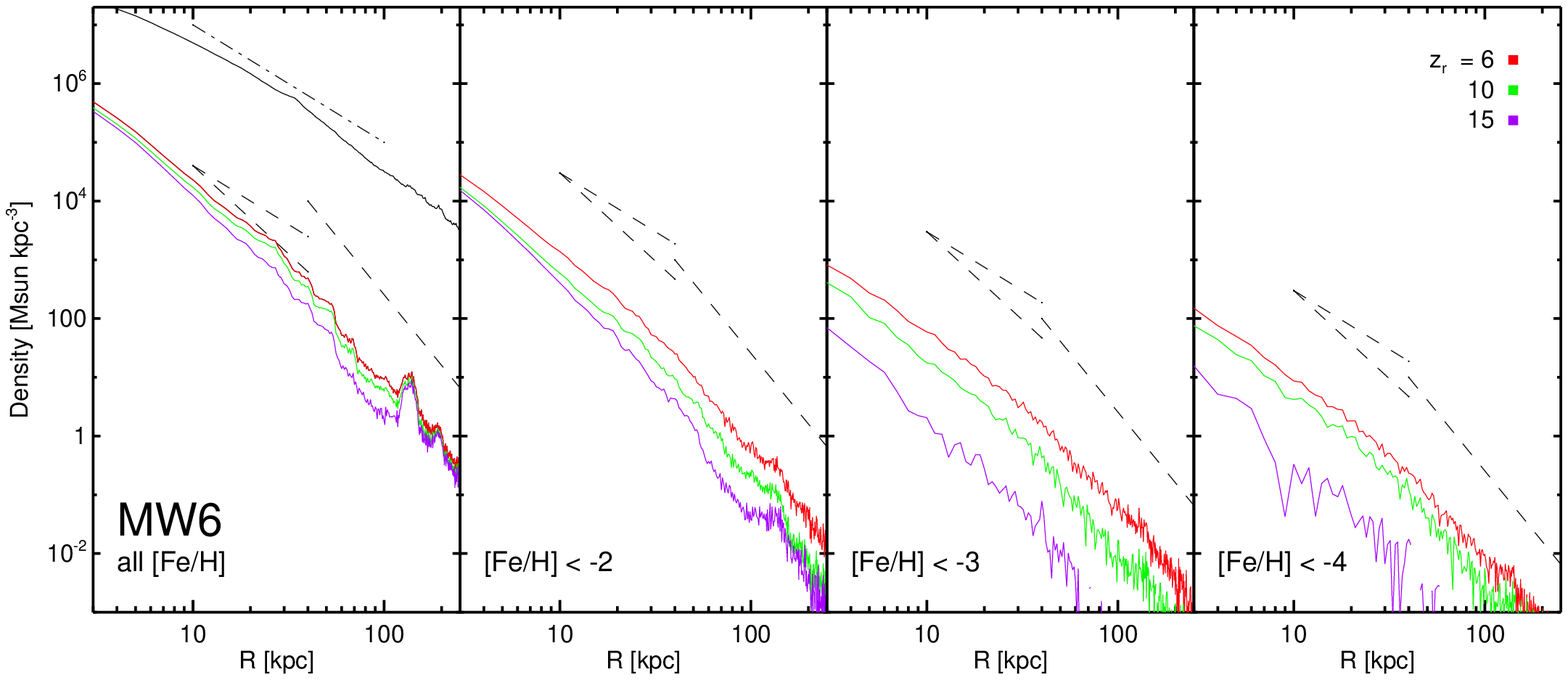}
\caption{Stellar mass density profiles for MW1 and MW6 for all stars (left panel) and three metallicity cuts, [Fe/H] $< -2$, $-3$, and $-4$ from left to right, and three redshifts of reionization in each panel: $z_r = 6.5$ (top), $10.5$ (middle), and $14.5$ (bottom). The dashed lines mark power laws with slope $-2$, $-3$, and $-4$.\label{densfig} }
\end{figure*}

The merger-tree-based chemical evolution models, with even simple prescriptions to describe the star formation histories of halo progenitors, give model halos that reproduce the observable properties of the real Milky Way stellar halo and its dwarf satellites to a reasonable degree. Figures~\ref{hmassfig} to \ref{mzfig} show the results of these tests in terms of the total stellar mass, mass density profile, and dwarf satellite properties.

First, the total stellar masses of the model halos agree well with observations and with other theoretical studies. Using star counts for millions of main sequence turnoff (MSTO) stars from SDSS, \cite{Bell:08:295} found a halo mass of $M_h = (3.7 \pm 1.2) \times 10^8$ \msun\ from $1 - 40$ kpc. \citeauthor{Bell:08:295} calculated the halo mass using star counts and an observationally-derived ratio of 4.7 \msun\ per MSTO star. This ratio corrects for the fact that only part of the stellar mass function is being observed by estimating the total present-day mass of a stellar population that contains one MSTO star (in this case, with color $0.2 < g-r <0.4$ measured from SDSS photometry). This ratio was determined empirically from observations of the globular cluster Pal 5. This observed ratio, which \citeauthor{Bell:08:295} note may be a lower limit, differs from the ratio 9.5 \msun\ per MSTO star that is correct for the IMF assumed here. Therefore the model halos  are compared to the \citeauthor{Bell:08:295} result corrected upwards by a ratio 9.5 / 4.7 and find excellent agreement between the model and the data. The total mass of the stellar halo is sensitive to the baryon accretion parameter $f_{bary}$ and to the star formation efficiency parameter $\epsilon _*$.  It is satisfying that the fiducial values chosen yield reasonable agreement with the data without any fine-tuning. 

Figure~\ref{hmassfig} also plots the total masses of the model halos ($R \leq 40$ kpc) for three cuts on metallicity. The total mass budgets of stars with low metallicity are reduced more by reionization than stars at higher metallicity. This effect is caused by the suppression of low-metallicity star formation in halos that are too small to accrete gas after reionization. This same effect can be seen as a steepening in the slope of the density profile as $z_r$ increases (Figure~\ref{densfig}). The significance of this effect for the origin of the inner stellar halo will be discussed further below.

The stellar mass density profile is another useful check from observations. This comparison appears in Figure~\ref{densfig} for MW1 (top) and MW6 (bottom). The four panels in each row show the stellar mass density profile for three metallicity cuts, and within each panel models for three different redshifts of reionization are shown. For comparison, power laws with slope $-2$, $-3$, and $-4$ are marked with dashed lines. The stellar mass density profiles including stars of all metallicity clearly fall in between slopes of $-2$ and $-3$ with a break apparent at $40-50$ kpc, in agreement with recent observations \citep{Morrison:00:2254, Yanny:00:825, Bell:08:295}.  Reasonable stellar masses and similar density profiles were also obtained by \cite{Bullock:05:931}, even though their modeling did not include a ``live" halo. Though observed density profiles are not available for the lowest metallicities owning to small numbers of known stars at [Fe/H] $\lesssim -3$, their overall normalization is consistent with the global metallicity distribution function (see \S~5.3). Alternative density profiles were calculated assuming lower values for the minimum halo circular velocity, $v_c^{min}$, that can accrete baryons after reionization, down to 15 \kms. These tests indicate that the results for the density / metallicity structure of the halo inside $R \sim 40$ kpc changes very little with lower thresholds; the main progenitor galaxies of the stellar halo are already above $v_c = 20 - 30$ \kms\ at reionization. There is a modest gain of order 50\% in the number of low-metallcity stars beyond $30-40$ kpc, for models with $v_c^{min} = 15 - 20$ \kms, which describes low-mass subhalos that form stars predominantly after reionization, if allowed, and are disrupted to form the outer halo. It appears that later tests for the halo within $40$ kpc do not depend on uncertainties in the baryon assignment prescription. 

\subsection{The Milky Way Satellite Luminosity and Metallicity Distributions}

The Milky Way's dwarf satellite galaxies have emerged as key indicators of the baryonic gas physics that govern star formation within small dark-matter halos. Their importance to the present model lies in their ability to serve as observational checks on the assignment of baryons to small halos and on the star formation histories of small halos under the influence of reionization. Recent discoveries of many faint dwarf galaxies by the Sloan Digital Sky Survey have now provided a measure of the dwarf galaxy luminosity function down to $M_V \sim -3$ \citep{Tollerud:08:277, Koposov:08:279}. Their properties are crucial to constructing a realistic model of the Milky Way, since the dwarfs provide information about galaxy formation on mass scales similar to those of the early Milky Way, but of course the true Milky Way progenitors are disrupted and mixed while the dwarf galaxies retain some of their stellar populations. 
Some of the modeling results in this section are already known to the literature on the subject; they are repeated here to check that the basic model is sound by comparing it to observations and previous theoretical results before moving on to perform novel tests of the origin and distribution of the Galaxy's oldest, most metal-poor populations. 

\begin{figure}[!t]
\begin{center}
\plotone{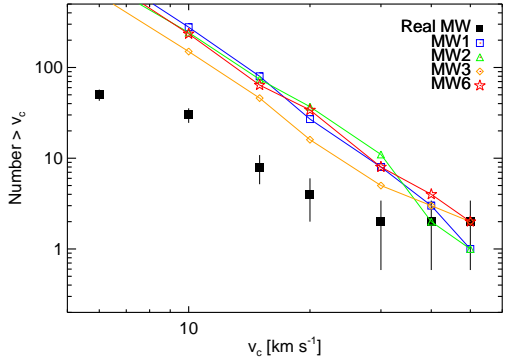}
\caption{Circular velocity distribution of substructure in the four complete halos, MW1, MW2, MW3, and MW6. The ``missing satellites'' problem is immediately apparent. \label{vc_hist_plot}} 
\end{center} \end{figure}

The central problem posed by the Milky Way's dwarf population is that it is much smaller than the expected number of subhalos in the host halo of the Milky Way's total mass; this is the well-known ``missing satellites'' problem \cite{Bullock:00:539}. This effect is shown in Figure~\ref{vc_hist_plot}, which shows the cumulative distribution of subhalo $v_c$ each in simulated host halo compared to the real Milky Way distribution drawn from \cite{Simon:07:313}. Note that the halo MW3, with slightly lower $M_{200}$ than the others, appears to better fit the real Milky Way system at $v_c \geq 30$ \kms. The relations and parameters chosen to model the hierarchical chemical evolution of the host halo should be able to explain the observed properties of the MW dwarf population in terms of their stellar populations; for verification of this model we investigate two relevant properties of the Milky Way dwarf population; the luminosity function (LF) and the luminosity-metallicity (LZ) relation. The LF is sensitive to how baryons are assigned to dark matter halos (as described in \S~\ref{baryonassignmentsubsection}), to how efficiently they form stars, and to the effects to reionization on the assignment and/or removal of baryons from small halos. The LZ relation is sensitive to all these, and also to the prescription for chemical and kinematic feedback (\S~\ref{feedbacksubsection}).

\begin{figure}
\includegraphics[width=3.6in]{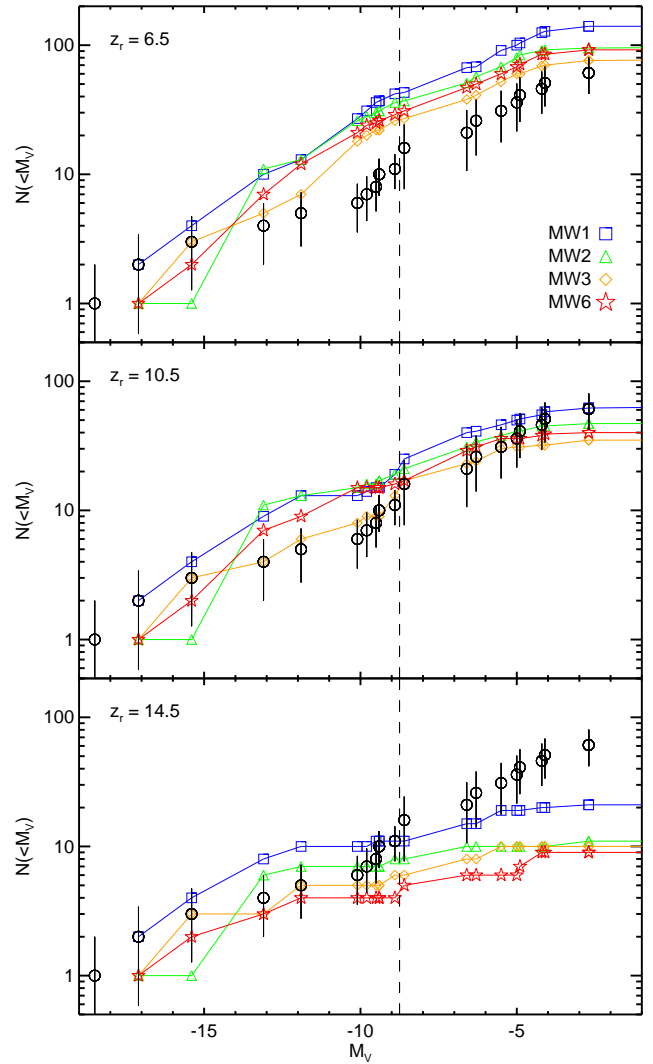}
\caption{Luminosity functions for the model dwarf satellite populations of the MW halos for three different redshifts 
of reionzation, $z_r = 7.5, 10.5$, and 14.5, top to bottom. 
\label{lumfuncfig}}
\end{figure}

Models for the Milky Way dwarf galaxy luminosity function appear in Figure~\ref{lumfuncfig} for three different redshifts of reionization. The observed luminosity function, marked in open circles, was constructed from the observed V-band absolute magnitudes and distances as compiled in Table 1 of \cite{Tollerud:08:277}. Since the comparison here is to all the dwarf galaxies within the present-day virial radius of the halo, Leo T at 417 kpc is excluded from this comparison. The SDSS-discovered dwarf galaxies are counted 5 times each to account for the fact that SDSS has covered only one-fifth of the sky. Together with the 11 classical dwarfs this weighted total of 50 gives 61 dwarf galaxies brighter than $M_V = -2.7$, with the cumulative total marked with $1\sigma$ errors from counting statistics marked in the figure. The model LFs are corrected for the difficulty of detecting faint dwarfs at large distance by excluding from the model count any dwarf subhalo that lies outside the SDSS completeness radius \citep[Equation 2]{Tollerud:08:277} at its magnitude. Though this simple correction is too sharp to be perfectly realistic \citep[cf.][]{Koposov:08:279}, it adequately captures the basic effect of finite detection limits in the observational surveys.

The fiducial values of $f_{bary} = 0.05$ and $\epsilon _* = 10^{-10}$ provide very good descriptions of the halo mass and density profile, so they are not varied here. To first order, the luminosity of a given dwarf depends linearly on their product, and so the need to have a few dwarf galaxies with $M_v \lesssim -15$ in the model halo to match the real MW means this product cannot be much lower than the fiducial value.

The next most important effect controlling the LFs of dwarf galaxies is the suppression of star formation in small halos by reionization, as illustrated in Figure~\ref{lumfuncfig} for $z_r = 6.5$, $10.5$, and $14.5$. It is apparent from the figure that the count of detectable dwarf satellite galaxies fainter than $M_V \sim -8$ declines rapidly as reionization moves to earlier redshifts, while brighter dwarfs are only little affected by earlier reionization. The reason for this behavior can be found in an analysis of the dark-matter assembly histories of the subhalos that host dwarf satellites with respect to the thresholds in halo circular velocity that enter the baryon assignment prescription. Recall from \S~\ref{baryonassignmentsubsection} that all halos are permitted to accrete their $f_{bary}$ share of baryons and so form stars prior to $z_{r}$, while only halos with $v_{c} \geq 30$ \kms\ can continue to accrete baryons after reionization. These latter halos will be relatively unaffected by the suppression of baryon accretion that reionization imposes, but those halos with $v_{c} \lesssim 30$ \kms\ will be limited in the total amount of star formation they can ever have by the time available to accrete gas prior to $z_{reion}$. It is these smaller halos whose total gas budgets are limited by $z_r$ that drop out of the visibility corrected LF for models with earlier reionization. Since the four model halos have different amounts of DM substructure, the best-fitting redshift of reionization varies but typically has $z_r \simeq 8-11$, with $z_r = 10.5$ adopted as the fiducial value for later comparisons. These basic results are in agreement with other recent attempts to explain the luminosity function of MW satellites in semi-analytic models (Koposov et al. 2009; Maccio et al. 2009).

\begin{figure}[!t]
\plotone{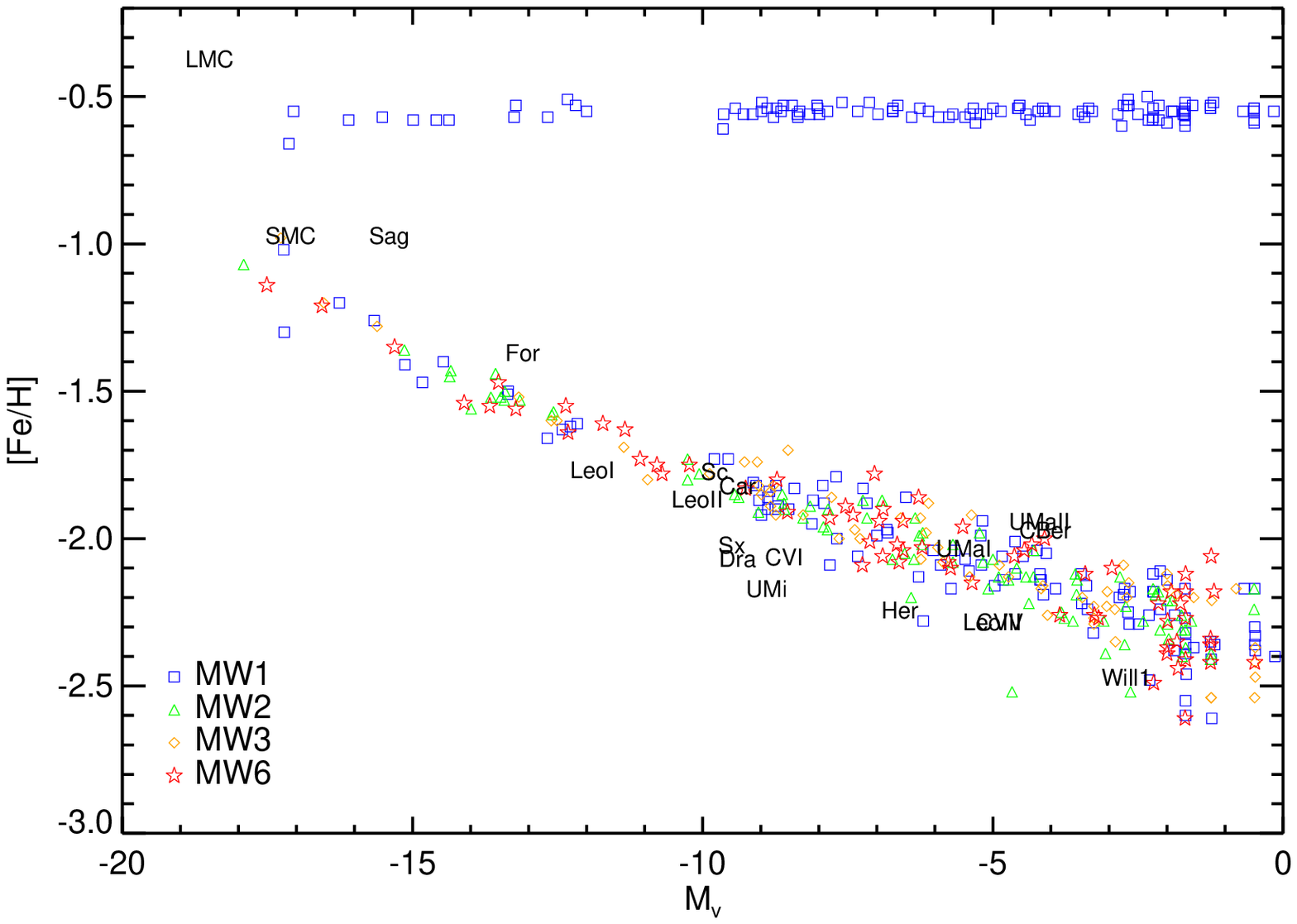}
\caption{The luminosity-metallicity relation for the MW halo and the four model halos. A single no-feedback model is shown for MW1. Models for MW1,2,3, and 6 are shown with $z_r = 10.5$ and $f_{esc}^Z = 50$. The slope of the relation and the increased scatter at low $M_V$ are both recovered in the models. \label{mzfig} }
\vspace{0.1in}
\end{figure}

The luminosity-metallicity (LZ) relation for the Milky Way dwarf population is shown in Figure~\ref{mzfig}, compared to the four model halos. The model metallicities are derived by obtaining a model MDF from particle-tagged stellar populations and taking the mean value. A model for MW1 with no SN-driven kinematic feedback (\S~\ref{feedbacksubsection}) is shown in open circles; all other models have $f_{esc}^Z = 50$, which is used as the fiducial value. As shown in similar modeling efforts \citep{Dekel:03:1131, Robertson:05:872}, models including selective feedback from small halos provide a good match to the overall trend in the real Milky Way. More information about the formation of the satellites and their role in building the stellar halo is available from theoretical modeling of chemical abundances; see in particular the series of papers by \cite{Font:06:585, Font:06:886, Font:08:215} and \cite{Johnston:08:936}.

The increasing scatter at low [Fe/H] is a real effect in the models arising from both the fully hierarchical treatment of structure formation and the stochastic treatment of chemical evolution. First, some of this scatter results from diversity of their merger tree histories, since subhalos that spend more time as small pieces prior to mergers will have lower [Fe/H] and $v_c$ at $z = 0$. The baryon accretion prescription also increases the scatter, since some halos have histories that extend longer than others before reionization, and some pop just above the gas accretion threshold after reionization and so can form stars later than those that do not. More scatter is added by the time variation of supernova rates from a stochastically sampled IMF, which inject metals into the gas reservoir and eject metals from the halo with a prescription (\S~\ref{feedbacksubsection}) that depends on the instantaneous supernova rate and the instantaneous halo mass (through the halo binding energy). All these effects contribute to increasing the scatter in the LZ relation at low luminosity, which appears roughly consistent with scatter in the observed metallicities of the known satellites. However, the data as they stand now are not sufficient to say whether all of these mechanisms acted in the real galaxies or which, if any were dominant.

\subsection{The Milky Way Halo Metallicity Distribution}

\begin{figure}[ht]
\includegraphics[width=3.5in]{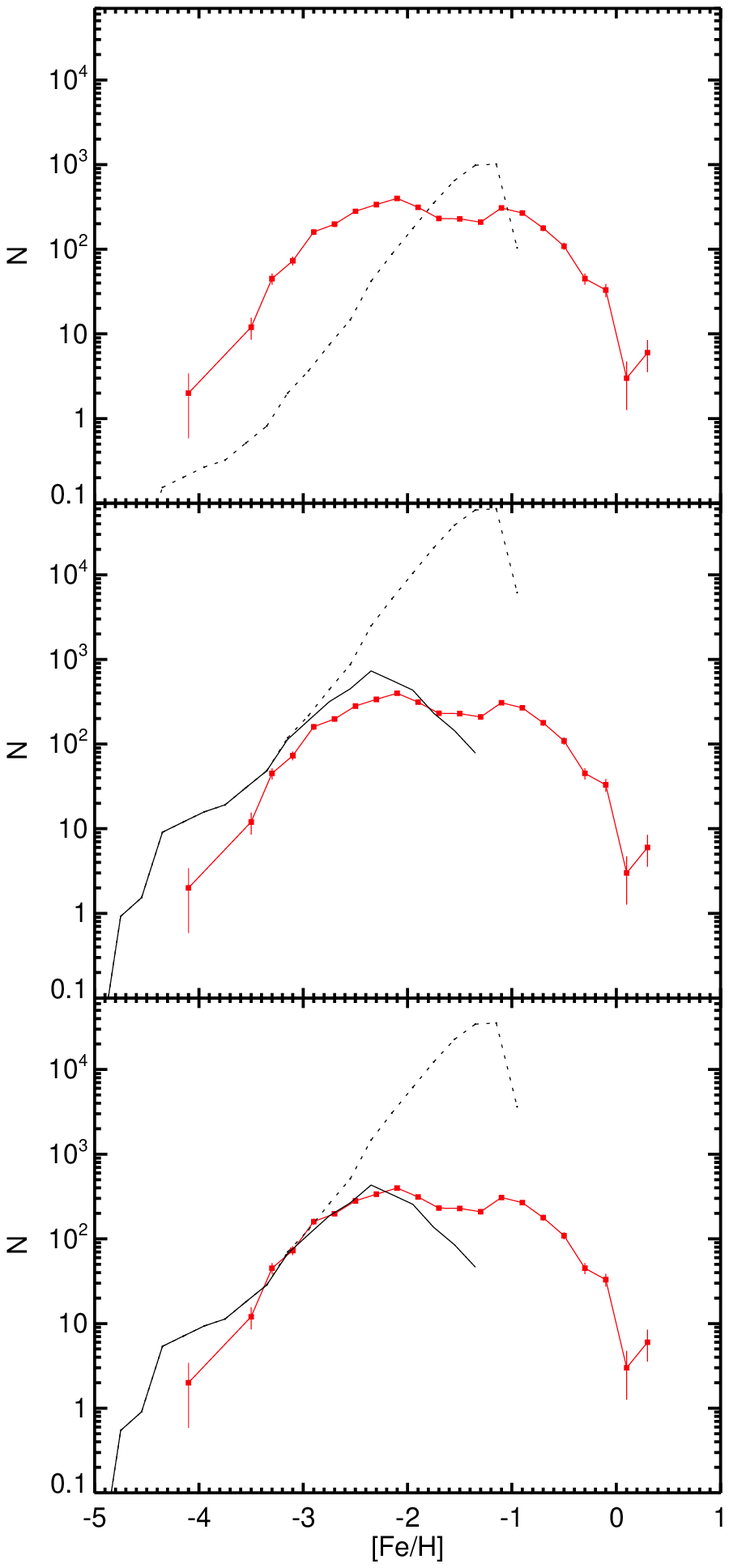}
\caption{Model MDFs compared to the HES MDF from \cite{Schoerck:08:1172}. At top, the observed HES MDF compared with the $R < 20$ kpc MW1 MDF normalized to a total of $N=3439$ stars at all metallicity. In the center panel, the model MDF has been corrected for the average HES visibility function and normalized as before. In the lower panel, the visibility correction is applied as before, but here the model MDFs are normalized to have 1507 stars at [Fe/H] $< -2$. In the two lower panels the uncorrected MDFs are marked with dotted lines. Except for the normalization, the dashed (uncorrected) and solid (corrected) curves are the same in all panels.  It is apparent that normalizing the model and observed MDFs where the visibility correction is smaller than $\sim 10$ or less leads to reasonable agreement. \label{mdffig} }
\end{figure} 

The Galactic halo metallicity distribution function (MDF) can be used to test the model halos, but it turns out to be a problematic comparison because of biases in the observed MDFs and uncertainties in the model. The MDF comparison is shown in Figure~\ref{mdffig}, where the MDF measured by the Hamburg/ESO survey (HES) of metal-poor stars is indicated in filled squares. Here the model is the particle-tagged MDF for all particles in the host halo within $R = 20$ kpc of the center. In the top panel, the HES MDF is shown with the exact number counts listed in Table 3 of \cite{Schoerck:08:1172} and error bars from Poisson statistics. Here the model MDF is renormalized to contain the same total number of stars ($N=3439$) as in the HES MDF. With no correction to either data or model, the agreement is very poor; the model MDF is too sharply peaked at [Fe/H] $\simeq -1.3$ and has too steep a tail to low metallicity. 

The HES was designed to select for EMP stars ([Fe/H] $<-3$), so \cite{Schoerck:08:1172} derive a correction function that accounts for the high degree of incompleteness at [Fe/H] $\gtrsim -2$ in the as-observed HES sample. This correction factor must be applied to the model MDF before the model and data can be compared directly. This comparison is performed in the middle panel of Figure~\ref{mdffig}, where the corrected model MDF now appears in the solid line and the uncorrected model appears in the dotted line (this is the same as in the top panel). Here the agreement between model and data is improved by the correction for the bias against stars with [Fe/H] $>-3$, but the fit is still unacceptably poor when the model and data are normalized to contain the same total number of stars. Given the very large magnitude of the visibility correction (a factor of 100 at [Fe/H] $=-1.7$, increasing at higher metallicity), it is perhaps wise to normalize the data to model in a region where the correction is minimized; that is, where the HES MDF is reliable {\it as observed}. This comparison appears in the final panel of the figure, where the agreement in the region where the comparison is meaningful is much improved. 

There is still some disagreement remaining at [Fe/H] $>-2$, but it is difficult to assess both the theoretical MDF and the observational MDF at these high metallicities. First, the visibility correction for the HES MDF becomes larger than a factor of 100 here, with presumably a large uncertainty in this correction. Second, above [Fe/H] $\sim -1.5$ the theoretical MDF becomes sensitive to the presence of star formation in the host halo (the main ``trunk'' of the tree) at redshifts of $z \sim 3$ and below. The exact star formation history here is uncertain since it involves the formation of the Galactic thick and thin disks, and it is unclear where and how the transition between disk and halo should be managed in the model. Star formation parcels that occur in the host halo after the last major merger are considered to be formed in the disk are are not included in any of the stellar population modeling. If in fact ``halo'' stars continue to form in the host after this milestone, then the halo MDF can be more populated with high [Fe/H] stars. Conversely, a visibility correction that is too large can produce disagreement in the sense seen here. Without exact knowledge of how HES stars are distributed kinematically into the halo or disk, and without a more detailed model of disk formation in the theoretical MDF, no more exact results can be obtained. The agreement seen between the two MDFs at [Fe/H] $\lesssim -2$ is, however, encouraging.


\section{Results III: Spatial, Kinematic, and Chemical Properties of Early Halo Stellar Populations}
\label{section-results3}

\subsection{Halo Chemical Evolution Histories} 

\begin{figure*}[!t]
\plotone{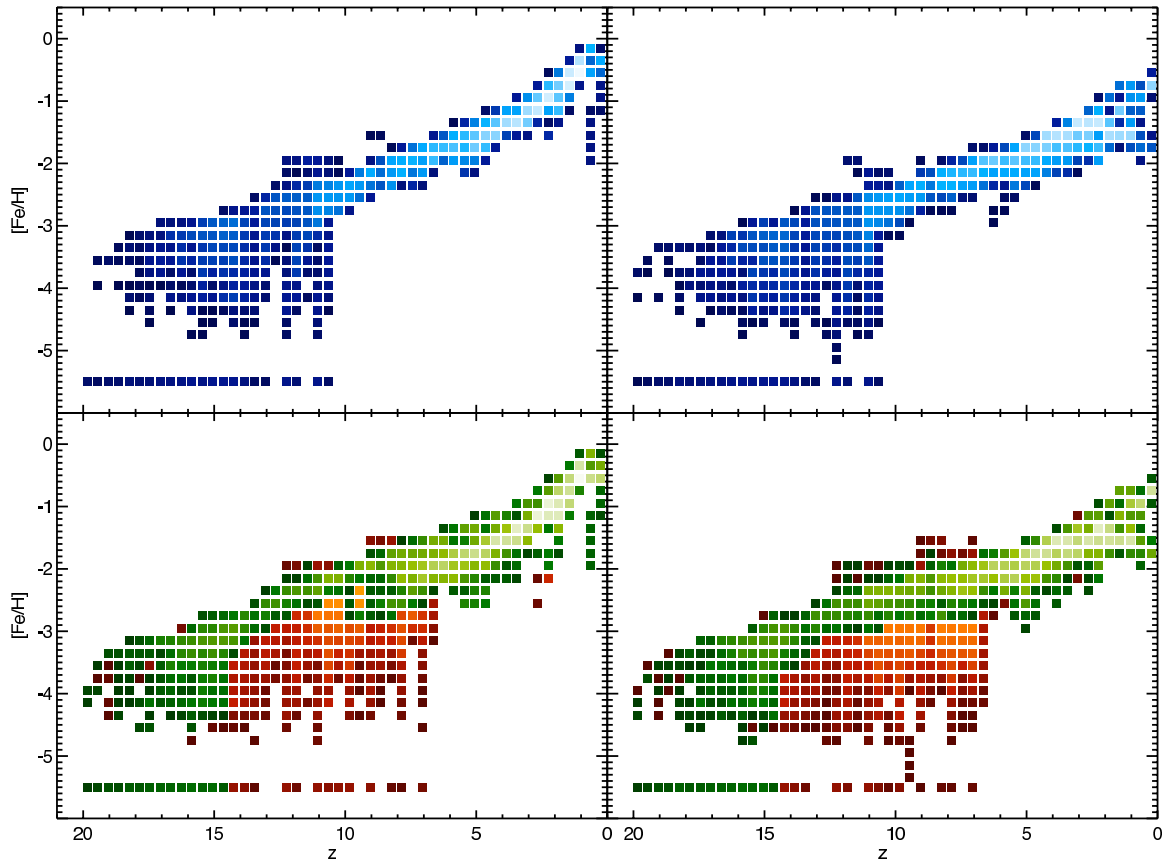}
\caption{Six chemical evolution histories showing the effect of chemical feedback and reionization. The left two panels have $f_{esc}^z = 0$ (no chemical feedback), the right two panels have $f_{esc} = 50$. The upper panels have $z_r = 10.5$, the lower two panels show $z_r = 6.5$ (red colormap) and $z_r = 14.5$ (green colormap). The effect of suppressing of star formation in small subhalos that would otherwise form low metallicity stars can be seen in the lower panels, where many redshift / metallicity bins are populated when $z_r = 6.5$ but not when $z_r = 14.5$. For this plot, star parcels with primordial abundances are displayed at [Fe/H] =$-5.5$.
\label{chemevfig}} 
\end{figure*}

A full halo model includes a self-consistent chemical enrichment history for the subhalos that end up in the host, and the particles that make up that halo. Example chemical enrichment histories appear in Figure~\ref{chemevfig}, which displays two-dimensional histograms of the chemical enrichment histories for six variations on MW1, with three redshifts of reionization with and without feedback. Several features of these distributions deserve comment. 

First, the hierarchical nature of halo assembly entails that chemical enrichment is inhomogeneous at a single time; halos of different mass form stars of different metallicity. This basic fact is observed in the Milky Way system today, where the disk forms stars at $\sim Z_{\odot}$ and the SMC forms stars at $0.1Z_{\odot}$. These models are of course constrained to achieve the current dwarf-satellite LZ relation. At some times, the range in metallicities is up to 3 orders of magnitude. 

Second, and perhaps more important for the goal of using low-metallicity stars to probe the high-redshift Universe, stars at a given metallicity form over a wide range of redshift. For instance, in the upper right panel where $z_r = 10.5$ and chemical feedback is included, stars with [Fe/H] $\sim -2.5$ form from $z = 13 - 5$. In a model where the halo at any early time consists of many smaller pieces, and the typical metallicity evolution is influenced by the mass scale, there must be such a spread in the formation times for stars at a given metallicity. These two trends -- a spread in metallicity at a single time and a range of times at a given metallicity -- 
are features that emerge from a fully hierarchical, non-homogeneous treatment of chemical evolution. Taken to their extremes, they would completely eliminate any correlation between metallicity and age and/or redshift; instead, in realistic models these two trends perturb the homogeneous picture where metallicity rises monotonically with time, but there is still a generally upward trend in metallicity over time.

Third, the suppression of baryon accretion and star formation in small halos after reionization affects the formation history of the most metal-poor stars. At any redshift, the relationship imposed between halo mass and chemical feedback (\S~4.2.7) ensures that the lowest metallicity stars are forming in the smallest halos, and these are the halos most affected by reionization. In the models displayed in the lower panels of Figure~\ref{chemevfig}, where $z_r = 6.5$, many redshift-metallicity bins are populated where they are not in models with an earlier reionization. This behavior, if it holds true in the real Galaxy, will have important effects on the use of these low-metallicity stars to probe the first stars. Of course, these star formation and metallicity histories are not readily compared directly to observation, and direct measures of stellar age generally are not effective for more than 10 Gyr before the present. We must therefore seek effective proxies for the age of the earliest populations, which is taken up next.

\begin{figure*}[!ht]
\begin{center}
\plotone{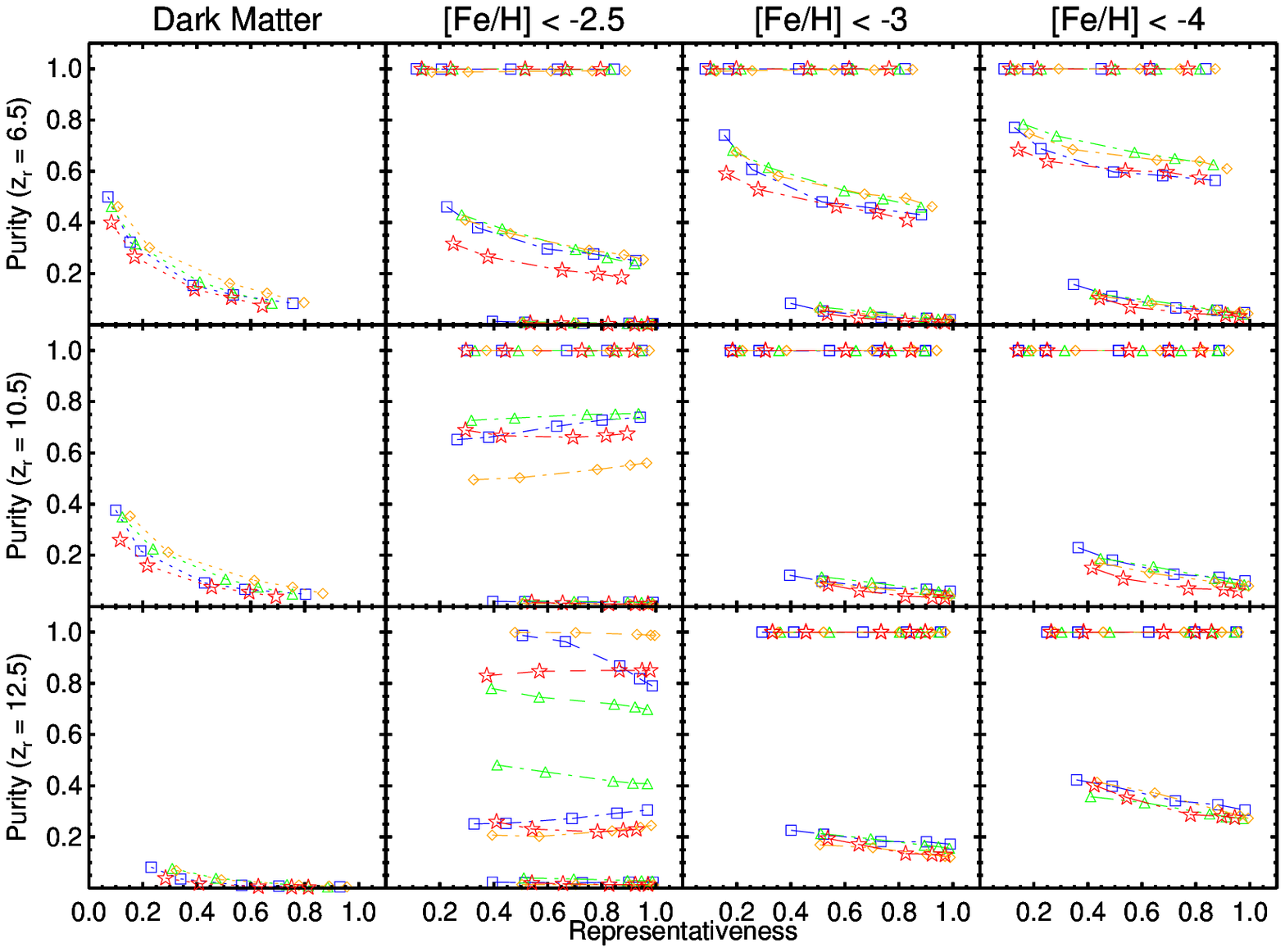}
\caption{Plot showing tension between obtaining a {\it distilled} sample of high-redshift structure and a {\it representative} sample of high-redshift structure. The y-axis measures the fraction of dark matter or stars below some metallicity and within some radius R that formed before some redshift. The three columns make three different metallicity cuts, [Fe/H] $< -2.5, -3$, and $-4$. The three rows mark three different redshift for reionization, $z_r  = 6.5, 10.5, 12.5$ (which has no effect on the dark matter). Within each panel, the curves show redshift cuts of $z_{cut} = 6, 10$ and $15$ from top to bottom. In several panels the curves for $z_{cut} = 6$ and 10 overlap at unity on the y axis, indicating that all stars of that metallicity formed before that $z_{cut}$. The irregular behavior observed for $z_r = 12.5$ and [Fe/H] $<-2.5$ results from few particles and poor statistics at that epoch and metallicity. \label{fracplot}} \end{center}
\vspace{0.1in}\end{figure*}

\subsection{Isolating and Sampling High-redshift Populations}

For Galactic Archaeology to succeed as a probe of star formation at early times, the metal-poor stars under study must in fact have formed at this early time. Of course, according to the ``chemical clock'' metal-poor stars are by definition more primitive. But the extended epoch over which, e.g. [Fe/H] $\sim -2$ stars form in the fiducial model suggests that stars of low metallicity may form well after reionization. Indeed, this same concern is raised by the discrepancies between the relative chemical abundances of the dwarf galaxies and the MW halo at the same metallicity \citep{Venn:04:1177}, indicating that the two Galactic components had different star formation histories leading to the same metallicity \citep{Font:06:585}. Thus, metallicity itself is at best an imperfect measure of chronological age in metal-poor stars, and additional evidence that the stars under study in fact probe the earliest times should be sought.

\begin{figure*}
\plotone{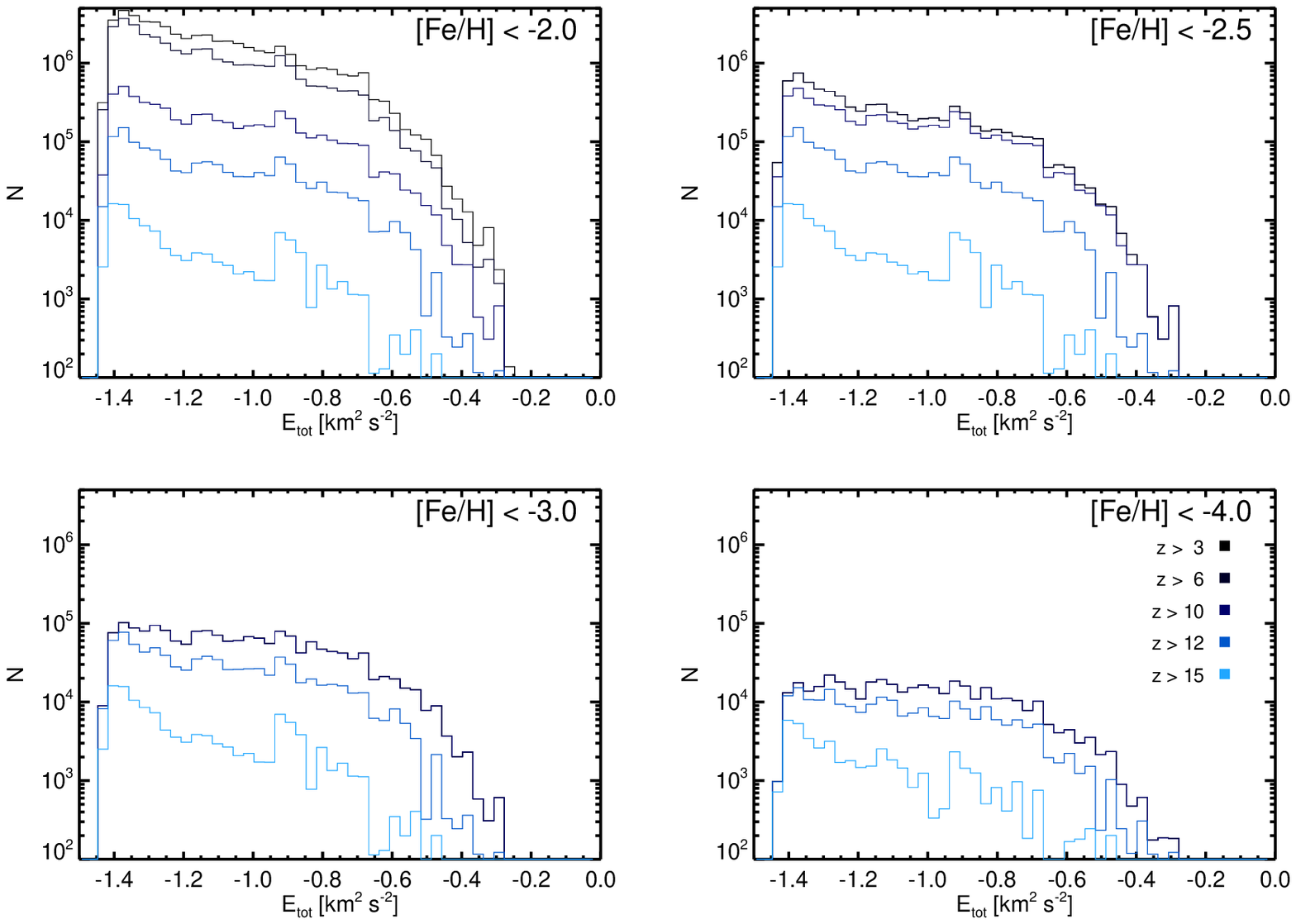}
\caption{The distribution of binding energy for stellar populations of various metallicities, cut by redshift of formation, for MW1. The histograms at upper left 
show all stars, with [Fe/H] $<-2.0$, and so on. In each panel, the contributions of stars formed before $z = 3$, 6, 10, 12, and 15 are marked. As traced by low-metallicity stellar populations, halos are clearly built from the inside-out, with the oldest stellar populations concentrated 
on the most tightly bound orbits with low $E_{tot}$. The regions with higher binding energy are filled in later. At the lowest metallicities, [Fe/H] $<-3$, all the surviving stars are in place by $z \sim 10$, since small halos where these would form are suppressed by reionization. \label{phasecutfig}} 
\vspace{0.1in}
\end{figure*}

Because the models are based on realistic dynamical simulations, they also express the distribution of formation time for stars of different metallicity from which we can estimate the fraction of stars at each metallicity that date from some high redshift. The interpretation of observational surveys involves two properties of a population: first, the fraction of stars at a given metallicity and within some volume that formed before some redshift $z_{cut}$, and second, how representative is that sample of all the stars in the halo that formed at that redshift. The first quantity, called ``purity'', is relevant because we would like to obtain a perfectly ``distilled'' sample of stars from high redshift, and this quantity measures how well this goal can be achieved (these two terms -- ``pure'' and ``distilled''  -- are used interchangeably here). The second is relevant because once a distilled sample is obtained it is desirable to know if it is a fair representation of all the stars that formed at that metallicity and at that redshift. Generically, one might expect that selecting for a pure sample will tend to reduce its use as a fair sample of its epoch, so that ``purity'' will correlate inversely with ``representativeness''.

Results from these tests are presented in Figure~\ref{fracplot}. The figure shows purity and representativeness for three different models that differ only in their redshift of reionzation $z_r = 6.5$ (first row), $z_r = 10.5$ (second row), and $z_r = 12.5$ (third row). The first column shows dark matter only; the latter three columns show the results for three different cuts on metallicity, [Fe/H]  $\leq -2.5, -3$, and $-4$ from left to right. Within each panel, there are three different redshift cuts displayed, $z_{cut} = 6, 10$, and 15 from top to bottom. The five points on each curve represent radial cuts of $R = 5, 10, 30, 50$, and 100 kpc. Several key results of interest emerge from this test. First, it is generically true that the model halos have a significant population of stars from before reionization, and that the fraction from before reionization increases at lower metallicity (that is, the curves shift up from one column to the next). Another notable feature is that the measure of purity increases at small radii - this is a consequence of the ``inside out'' construction of CDM halos, with inner regions formed from earlier subhalos. This behavior in the dark matter, at left, is preserved in the stars. Next, note that at a given radial cut the representativeness increases for higher $z_{cut}$; this is another effect of the central concentration of stars from the earliest halo progenitors. For example, [Fe/H] $< -3$ stars within 5 kpc of the halo center are approximately 50\% of all [Fe/H] $<-3$ stars from before $z = 15$, while they are only $\sim 15-20$\% of all [Fe/H] $< -3$ stars from before $z = 10$ (third column). In other words, the chemical clock is gets going earlier in the center of the halo than in the outer regions. Finally, it is notable that these results depend only a little on the redshift of reionization (that is, the curves shift only a little moving down the column as $z_r$ increases, moving more only when $z_{cut} \sim z_r$ ). Thus uncertainty in the redshift of reionization does not substantially undermine the goal of obtaining ``distilled'' samples of high-redshift stars. 

\begin{figure*}[!ht]
\plotone{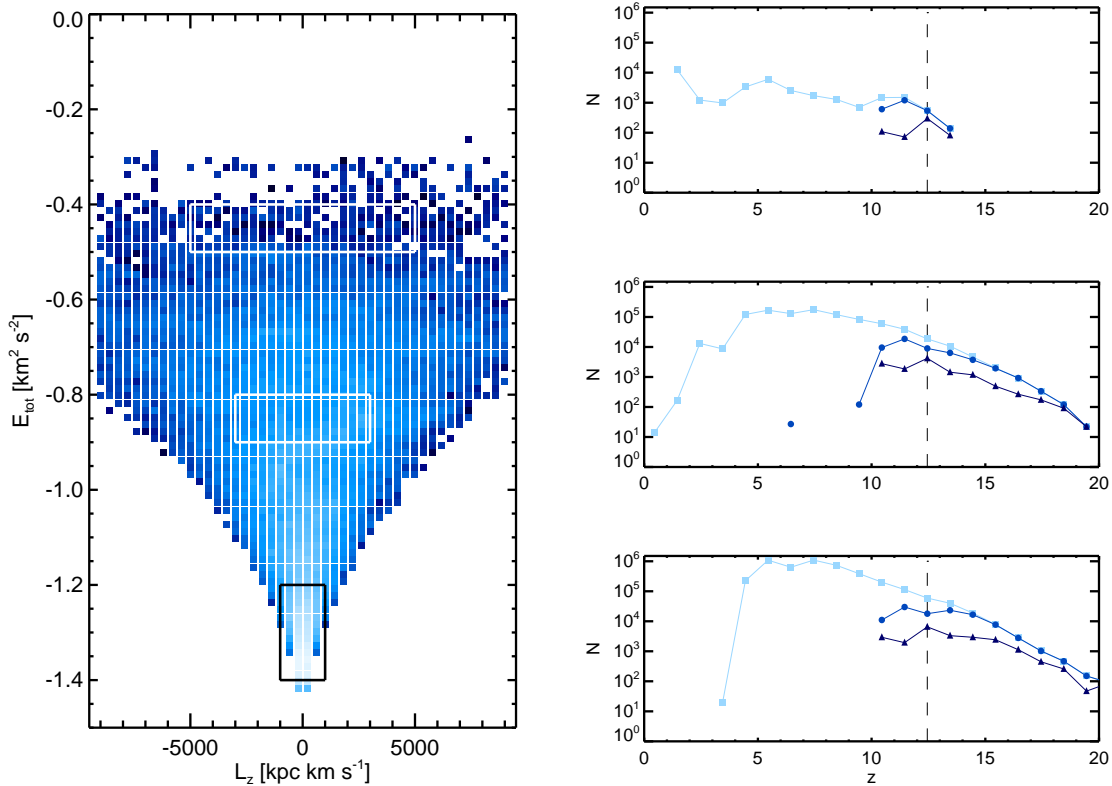}
\caption{At left, a 2D Lindblad histogram of all stars in MW1 with [Fe/H] $<-2$ and $R < 100$ kpc. Three regions of interest in $L_z - E_{tot}$ space are 
marked. Their particle-tagged star formation histories are shown at right for stars of [Fe/H] $<-2$ (filled squares), $<-3$ (filled circles), and $<-4$ (filled triangles).  For more tightly bound orbits (lower $E_{tot}$), the star formation histories extend to higher redshift.  Inspection of Figure~\ref{mhist-plot} shows that the increase in the highest redshift of star formation for stars on more tightly bound orbits reflects the underlying dark matter history. \label{mhistfig}} 
\vspace{0.1in}
\end{figure*}

If the earliest, most metal-poor stars in the Milky Way halo lie closer to the center of the halo than the typical metal-poor star, there may be kinematic signatures that could be exploited to further isolate them from confusing populations. Figures~\ref{phasecutfig} and \ref{mhistfig} show how the star formation histories of different halo components correlate with the total binding energy of the particles. Figure~\ref{phasecutfig} shows that, just as for the dark matter, stellar populations at low metallicity form and assemble from the ``inside-out''. This figure shows the distribution of binding energy for halo stellar populations cut at four different metallicities. The cumulative histograms in each panel show the distribution of stellar populations formed before five different redshifts: $z = 3$, 6, 10, 12, and 15. Note that in each panel, the highest-redshift cuts provide a larger fraction of stars on tightly bound orbits (low $E_{tot}$), and that the more loosely bound regions of the halo are filled in later. This is a generic consequence of the ``inside-out'' growth of dark matter halos, and the tendency of the earliest bound material to lie in the center of large halos at late times. Note also that the lowest-metallicity stars, [Fe/H] $<-3$, in the lower two panels, have all formed by $z = 10$, so that later curves lie on top of one another. Only stars at [Fe/H] $\gtrsim -2.5$ continue to form after $z \sim 10$. 

Figure~\ref{mhistfig} shows how the star formation histories associated with particles vary with the binding energy of those particles, as a function of redshift and metallicity. The star formation histories for the least bound particles ($E_{tot} \sim -4\times 10^{4}$) begin in small halos that form late compared with the earliest progenitors of the host, so their most metal-poor stars form at $z \sim 10-13$ and their histories end at reionization. Stars on more bound orbits ($E_{tot} \sim -8\times 10^{4}$) extend back to $z \sim 20$ and continue to form in significant numbers at [Fe/H] $\sim -2$ after $z = 5$. The most bound particles  ($E_{tot} \sim -1.3\times 10^{5}$) carry stars formed at $z > 20$ and have very few [Fe/H] $< -2$ stars formed at $z < 5$, when the host is forming stars of much higher metallicity (see \ref{chemevfig}). This is caused by the fact that most of these particles have been in the host or its closest merger partners since the very beginning, they formed their metal-poor stars very early, and have not had a merger that brought in metal-poor stars on such tightly bound orbits. Note that the lowest selected region in Figure~\ref{mhistfig} contains almost ten times as many stars from $z \sim 20$ as the middle region.  From these tests it is evident that simple model descriptions of the MW halo imply that significant populations of stars remain from the epoch of reionization, and that observational samples designed to select stars within the inner few kpc of the Galaxy and at low metallicity will have the best chance of finding the true survivors of reionization.

This test suggests that searches for metal-poor stars on the most tightly bound orbits that can be accessed with a given observing strategy will be able to find preferentially older metal-poor stars. Modern observational surveys can already kinematically select more bound populations using radial velocities, proper motions, and a model of the Galactic potential \citep{Carollo:07:450, Morrison:09:694}. A simple application of these results would use such a kinematically selected sample to study relative chemical abundances such as [$\alpha$/Fe] and $r$ and $s$-process abundances as a function of orbital binding energy for each star. To obtain useful leverage, it would be helpful to study populations within all three binding energy regions marked in Figure~\ref{mhistfig}, where the least bound region contains no stars from $z > 14$ and the most bound region includes no stars from $z < 3$. Of course, since all three regions contain large, metal-poor populations from $z \simeq 4 - 10$, any variation will likely occur in the tails of the distribution of abundances and it may be necessary to obtain significant samples of stars in all three regions to detect any differences in relative chemical abundances with binding energy. 

\section{Results IV: Implications and Applications for the Study of Metal-poor Stars} 
\label{section-results4}

The foregoing sections have described the detailed results of the halo-building models. This section summarizes the key conclusions of this study, and then draws implications for the further study of metal-poor halo stars and their use as probes of star formation in the early universe. 

\begin{itemize}
\item[1.] A simple model that takes into account only (a) baryon accretion, (b) star formation, (c) selective loss of metals from small halos, and (d) the effects of reionization on the IGM, can give an effective description of the gross properties of the Milky Way stellar halo: the stellar mass, density profile, metallicity distribution, and the luminosity function and luminosity-metallicity relation of the dwarf satellites (Figures 5 to 10). 
\item[2.] When the formation histories of Milky Way halo progenitor galaxies are summed together, two important trends emerge. First, stars at a given metallicity form over a long period, and second, at a given time there is a wide range of metallicity (Figure~\ref{chemevfig}). Thus the ``time clock'' and ``chemical clock'' are decoupled even over relatively small regions in the Milky Way halo, and detailed modeling is necessary to relate one to the other at any point in time or metallicity. 
\item[3.] Because of the inside-out construction of dark matter halos, stars at a given metallicity are older near the center of the halo when compared with stars at larger radii or on more loosely bound orbits. These inner regions are therefore more pure distillations of high-redshift populations, but are conversely less representative of all populations at that metallicity and redshift (Figure~\ref{fracplot}). Because reionization suppresses small, presumably low-metallicity halos at late times, large fractions of all [Fe/H] $\lesssim -2$ stars are from $ z> 6$, while $20-40\%$ of all stars of [Fe/H] $=-4$ to $-3$ formed at $z \gtrsim 10-15$. 
\item[4.]  These last two effects combine for the trend seen in Figures~\ref{phasecutfig} and \ref{mhistfig}: as binding energy with respect to the Galaxy's potential well increases (as stars become more tightly bound), the mean redshift of star formation and the earliest redshift of star formation both shift to earlier times. For example, the fraction of $\lesssim -2$ stars from $z > 15$ increases. Weakly bound halo populations ($R > 40$ kpc) may have very few if any stars from $z \gtrsim 12$ (except in subhalos). Because of general chemical evolution and suppression of late, low-metallicity star formation by reionization this preference  for high-redshift stars on tightly bound orbits is stronger in the stars than in the dark-matter alone.  
\item[5.] Because old, metal-poor stars prefer tightly bound orbits but exist throughout the halo (that is, populate a large region of phase space even if they prefer one part of it), there is a natural tension between obtaining a {\it distilled} sample of high-$z$ stars and a {\it representative} (however defined) sample of high-$z$ stars. 
\end{itemize} 

\begin{figure}[!ht]
\plotone{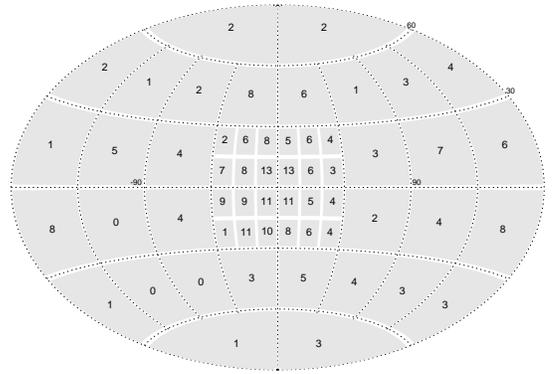}
\caption{The fraction of low-metallicity (EMP) stars from high redshift as a function of position on the sky. In each bin, the percentage of stars with [Fe/H] $\leq -3$ that formed prior to $z = 15$. The halo center of mass (``Galactic center'') lies at the zeropoint of latitude and longitude. \label{skymapfig}} 
\end{figure}

With the knowledge that old stellar populations are preferentially concentrated in the inner regions of the Galactic halo (Figures~\ref{fracplot}, \ref{phasecutfig}, and \ref{mhistfig}), it is now possible to calculate how these populations should be distributed on the sky as a first step to evaluating the ability of present or future observational campaigns to detect the chemical abundance signatures of the first stars. Figure~\ref{skymapfig} shows the variation of the ``purity'' of a sample of stars at low metallicity on the sky, [Fe/H] $< -3$ and $z \geq 15$. Because these high-redshift, low-metallicity stars are represented in the simulation by a relatively small number of particles, it is necessary to coadd the four model halos and average the probabilities for the resulting composite over relatively large areas of the sky to reduce numerical noise and artificial clustering. Nevertheless, it is clear that the halo populations with the highest fraction of high-redshift stars -- those most likely to carry the chemical abundance signatures of the first stars -- cluster near the Galactic center. These low-metallicity stars reside ``in the bulge'', though they are not ``of the bulge''. Near the Galactic center, EMP stars from $z > 15$ account for $10 - 15$\% of the population, while far away from the center toward the anticenter and poles they are typically less than $2-5$\% of the population. 

This test suggests that observational surveys may be able to separate the oldest EMPs from later populations, if they can efficiently select and study EMP stars in the crowded regions within $10-30^{\circ}$ of the Galactic center  (5 kpc subtends an angular region 30$^{\circ}$ at 8.5 kpc). While the the HK, Hamburg-ESO, and SDSS surveys have shown the EMPs exist all over the sky, not all EMPs are created equal - those near the Galactic center are generically older than those in the outer regions of the galaxy. And while ``low metallicity'' is equivalent to ``first'' on the chemical abundance clock, it is not necessarily so on the chronological clock. Since EMP stars do not otherwise reveal their time of formation, it is important to find some proxy for formation time that will enable correlations of chemical abundance signatures with time. With sufficiently large samples, it should be possible to use this variation on the sky as leverage to isolate the chemical abundances of the first stars in a differential comparison - any differences in the abundances of EMPs stars in the two regions could be caused by earlier time of formation for stars in the innermost Galactic halo. 

Stellar abundance surveys to date have tended to avoid the innermost Galaxy, where observations are hindered by the density of stars on the sky and high visual extinction. The HK, HES, and SDSS surveys all focused on high Galactic latitude for these practical reasons. Yet the stellar halo of the Milky Way extends all the way to the center of the potential well, and as shown above the oldest stars concentrate near the very center. While even surveys that avoid the bulge entirely should contain some small percentage of EMP stars from very high redshift, using the leverage provided by the ``inside-out'' growth of the halo will require obtaining an EMP sample that attempts to maximize the fraction of the very oldest EMP stars. The near-infrared APOGEE survey planned as part of SDSS-III\footnote{http://www.sdss.org} will be the first major effort to obtain a large set of stellar abundances near the center of the Galaxy, and as such it has a chance of detecting a difference in the abundance patterns of the innermost halo stars and those seen in SDSS/SEGUE, which avoids the bulge.

What sorts of questions might be addressed when observational surveys can probe the innermost, oldest stars in the Galaxy? One such question is motivated by recent studies of the carbon-enhanced metal-poor stars. The frequency of these CEMP stars has been linked to the IMF in the binary mass-transfer model for their abundances \citep{Lucatello:05:833, Tumlinson:07:1361}. If the early IMF is limited by the cosmic microwave background, as suggested by \citet{Tumlinson:07:L63}, it should have a strict time-dependence driven by the 1+$z$ dependence of the CMB temperature.  Thus the CEMP frequency should increase in regions of the Galaxy containing the oldest stars and decrease where the mean stellar age at any metallicity is lower. Given the results above, CEMP frequency should increase for stars with lower binding energy and for stars near the Galactic center. Such a change for stars with the same metallicity could provide key evidence that the CEMP phenomenon is linked to the CMB and the IMF. Similar arguments can be made for the chemical clock provided by the abundances of $r$-process elements in metal-poor stars~\citep{Barklem:05:129}. As \feh\ increases, the scatter in r-process abundances appears to decline as the star forming material in the early galaxy became more enriched in r-process elements, and more homogeneously mixed. If the more tightly bound inner halo stars formed earlier, they might be expected to show larger scatter in the r-process abundances than less tightly bound, and so later forming, stars at the same \feh. 

These considerations suggest that a new frontier is opening in the study of metal-poor stars and their use as probes of star formation near the end of the cosmological Dark Ages. Very large samples of EMP stars can now be obtained by surveying large areas of sky. Metallicities can be estimated for large fractions of these samples, but high-precision abundances can be measured for only some smaller subset. To answer the questions of interest, including those about the first stars, these samples must be selected carefully, with knowledge of how the selection relates to the underlying question and how it introduces biases. For judicious selection, it is important to have a full synthetic model such as the one attempted here.

This study has advanced but not fully met all the enumerated goals laid out in Section 1. It is clear that simple halo building models can successfully describe the bulk properties of the Milky Way stellar halo without excessively complex parameterizations (Goal 1). These ``vanilla'' halo models possess significant populations of stars dating from prior to reionization and allow the chemical abundances and spatial distribution of these stars to be quantified (Goal 3). Reionization and chemical feedback have been identified as key physical influences on the resulting ancient stellar populations (Goal 2), but other physical influences such as internal mixing within galaxies and local ionization effects have been left for future work. These models show that stellar populations can be ``distilled'' to prefer stars dating from reionization by selecting for low metallicity ([Fe/H] $\lesssim -2.5$) and for stars on tightly bound orbits in the center of the halo; these selection criteria have also been quantified (Goal 4). Yet much work remains to be done to refine and extend the framework on which observational tests in ``Galactic Archaeology'' will be performed. Many important physical ingredients can be studied with departures from the vanilla model; two important examples are non-local metal transport between neighboring halos and departures from a normal IMF. These tests will be performed in future studies, now that the behavior of simpler models is known. Finally, increased numerical and time resolution will allow the very smallest, highest redshift halos to be modeled self-consistently in the proper Galactic context. With a fully realized ``Virtual Galaxy'', and observational surveys that can advance into unexplored regions of the galaxy with ever-increasing fidelity, we can hope to find the residues of the first stars in the Milky Way itself.

\acknowledgements
I gratefully acknowledge the support of the Gilbert and Jaylee Mead Fellowship in Astrophysics in the Yale Center for Astronomy and Astrophysics, where a portion of this work was completed. Paolo Coppi donated computing time for the simulations, and the Yale ITS/HPC staff provided the necessary hardware and software support. I am also grateful to Romain Teyssier for the parallelized version of GRAFIC1. Thanks are also due for helpful comments from an anonymous referee. Support for a portion of this work has come from the Director's Discretionary Research Fund at STScI. 

\bibliographystyle{/Users/tumlinso/astronat/apj/apj}
\bibliography{/Users/tumlinso/astronat/apj/apj-jour,ms}

\end{document}